Fetal oxygen delivery and consumption and blood gases in relation to gestational age.

D.W. Rurak[1], M.Y. Shen[2] and K.S. Joseph[3]

[1,3]BC Children's Hospital Research Institute, Department of Obstetrics & Gynecology,

University of British Columbia, Vancouver, BC, Canada

[2]Department of Psychiatry, University of Calgary, Calgary, Alberta, Canada

Emails:

Dan Rurak, drurak@cw.bc.ca

Mary Shen, mary.yiting.shen@gmail.com

KS Joseph, ksjoseph@bcchr.ca

Corresponding Author and author to whom offprints should be sent:

Dan Rurak
BC Children's Hospital Research Institute
950 West 28th Avenue
Vancouver, BC, Canada
V5Z 4H4
drurak@cw.bc.ca

Funding support for this manuscript, particularly for the publication charge, come from a Canadian Institutes of Health Research grant held by KSJ.

**List of Abbreviations:** ACTH, adrenocorticotrophic hormone, AMP, adenosine monophosphate GA, gestational age; Qum, umbilical blood flow; $Do_2$, oxygen delivery; $Vo_2$, oxygen consumption; PlGF, placental growth factor; sFlt-1, soluble Fms-like tyrosine kinase receptor; Seng, soluble endoglin; VEGF, vascular endothelial growth factor; KDR, kinase-inserted domain receptor; REM, rapid eye movement

Abstract (250 words)

Oxygen crosses the placenta by diffusion and placental permeability to $O_2$ is high. Thus the fetus receives adequate amounts, but vascular $Po_2$ is much lower than after birth. Studies of sustained fetal hypoxemia and acute 40-45% hemorrhage show that hypoxemia is not tolerated whereas hemorrhage is. This suggests that if fetal $Po_2$ falls markedly, $O_2$ diffusion from blood to tissue is impaired. Uterine blood and umbilical blood flows/fetal weight fall progressively with advancing gestation. This results in fetal hypoxemia, an increase in $Pco_2$, and decrease in pH. This decreases fetal $O_2$ delivery, and in fetal lambs and horses there is a decrease in fetal $O_2$ consumption. The decrease in $O_2$ demands is linked to a decrease in fetal breathing and body movements and growth rate. The decrease in fetal motility is due to an increase in fetal plasma $PGE_2$ concentration, which begins at ~120 days GA in sheep and is due to the prepartum rise in fetal cortisol. Also, adenosine administration to fetal lambs decreases fetal breathing and REM sleep and the plasma adenosine concentration increases in late gestation. The fetal plasma levels of neurosteroids, which suppress fetal motility, increase with advancing gestation. The prepartum cortisol rise also inhibits fetal growth. In normal pregnancies, these mechanisms operate effectively to maintain an appropriate balance between fetal oxygen consumption and delivery. However, in pregnancies with either further reduce $O_2$ delivery or increase fetal $O_2$ demands, the mismatch between $O_2$ delivery and consumption may worsen leading to IUGR, hypoxic organ damage or stillbirth.

Note that the animal experiments conducted by one of the authors (DWR) and discussed in this review were approved by institutional Animal Care Committees and were consistent with the regulations and guideline of the Canadian Council on Animal Care.

1.0 Placental Oxygen Transfer

Maternal to fetal transfer of oxygen via the placenta occurs by simple diffusion. However, the venous drainage of the placenta on the maternal (uterine vein) and fetal (umbilical vein) sides of the placenta are relatively removed from the sites of oxygen transfer and receive blood from non-exchange areas. This has hindered attempts to obtain detailed information on placental $O_2$ transfer, but using techniques developed to study gas exchange in the adult lung, particularly the placental diffusion of carbon monoxide, which can be used to estimate the diffusing capacity of oxygen, useful information has been obtained. The placental diffusing capacity for carbon monoxide in the fetal lamb averaged 0.55±0.02 (SD) min.mmHg.min/kg fetal weight[1]. When normalized to fetal weight, this value is similar to that from the adult lung, but when normalized to lung weight the value from fetal lambs is less than that of the human and this is also the case with placental CO diffusing capacity determined in the fetal dog[2]. This suggests that in terms of $O_2$ transfer the placenta is less efficient than the adult lung.

1.1 Factors Affecting Placental Oxygen Transfer

Longo and coworkers[3] used a mathematical model to examine the effect of varying the individual variables in their model on placental $O_2$ transfer. Of the variables examined umbilical arterial $Po_2$ and fetal blood affinity for $O_2$ had the largest effects followed by maternal and fetal placental blood flow. In contrast, alteration in the diffusing capacity for $O_2$, particularly increases in the capacity had minimal effects. The importance of umbilical arterial $Po_2$ is explained by the fact that it is one of the determinants of the maternal-fetal $O_2$ gradient. Hill et al[4] used their estimate of the diffusion capacity of $O_2$, along with maternal and fetal inflowing $O_2$ tensions and maternal and fetal placental blood flows to develop a mathematical model of placental $O_2$ transfer. The model predicts $O_2$ equilibration between maternal and fetal blood in the placental exchange units.

The finding of Longo et al[3] that alterations in maternal and fetal blood flows have a major impact on placental $O_2$ transfer is similar to the research conducted by Meschia and Battaglia on diffusion and flow limited placental clearance[5,6]. The basic experimental approach is to infuse compounds intravenously to fetal lambs until steady-state conditions exist. At that point, assuming that the fetus can neither metabolize nor excrete the compound via its kidneys, the compound is being transferred to the mother via the placenta and the placental clearance of the compound is equal to the infusion rate. Molecules that are hydrophilic, polar, charged or large have a low placental clearance and permeability and their placental transfer is largely determined by this. They are diffusion limited. There is not equilibration between maternal and fetal blood in the placenta and changes in maternal or fetal placental blood flows have minimal effects on placental transfer. In contrast, small hydrophobic compounds have high permeability and changes in permeability, particularly increases, have negligible effects on placental transfer. These compounds are affected by changes and maternal or fetal placental blood flows and are thus termed flow limited in terms of placental clearance. The ratio of maternal/fetal volume flow is also important with placental clearance being maximal when the ratio is $1^5$.

1.2 Orientation of Maternal and Fetal Placental Blood Flows

In addition to the volume flows of maternal and fetal placental perfusion and the ratio between maternal and feta blood flow, the orientation of maternal and fetal placental blood flows is also important of the transfer of highly diffusible compounds. Histological and functional studies in several species have shown a number of different arrangements of the maternal and fetal placental vasculatures. These include a counter-current arrangement, which is found in the mouse[7], guinea pig and horse[8]. In the sheep, several orientations have been proposed including counter-current and cross-current[9]. However, Wilkening and Meschia[10] have concluded that functionally the sheep placenta behaves as a venous equilibrator, that is there is equilibration between maternal and fetal placental venous flows. There have been estimates of the efficiency of the different orientations, using equations that were initially developed to assess heat exchangers (see [11]). With the $O_2$ diffusion capacity of the placental divided by the fetal transport capacity for $O_2$ on the x axis and fetal arterial $Po_2$ plotted as a percentage of maternal arterial $Po_2$ on the y axis, the analysis predicts that with a counter-current orientation fetal umbilical venous $Po_2$ would be ~70-80% of maternal arterial $Po_2$, with a multivillous orientation ~60% and with a concurrent exchanger ~50%.

1.3 Fetal Vascular $Po_2$

Table 1 gives $Po_2$ values in the maternal and fetal vessels supplying the placenta in those species from which value are available. $Po_2$ is the fetus is much lower than predicted in the papers cited above[12]. In no species is the value in the umbilical vein higher than in the uterine vein or intervillous space.

Table 1. Representative values for Po2 (mm Hg) in maternal artery (MA), uterine vein (UtV) umbilical vein (UV) and umbilical artery (UA). Also given in the difference between the Po2 from UtV and UV. (modified from Rurak[12].

| Species | MA | UtV | UV | UA | UtV - UV |
|---|---|---|---|---|---|
| Sheep | 95 | 57 | 35 | 20 | 22 |
| Cow | 98 | 59 | 38 | 26 | 21 |
| Pig | 84 | 52 | 31 | 21 | 21 |
| Horse | 95 | 50 | 49 | 33 | 1 |
| Human | 95 | 50* | 41 | 29 | 9 |
| Guinea Pig | 114 | | | 30 | |

*, intervillous space. Data from Silver & Comline[8], Faber and Thornburg[25], Soothill et al[80], Bjellin et al[71].

The postulated reasons for this discrepancy between the predicted and measured values include the low placental diffusing capacity of oxygen when normalized to placental weight, the mixing of maternal and fetal placental venous flow with blood for other regions, umbilical diffusion shunting of oxygen and other highly diffusible compounds[10] (transfer of the compounds for fetal venous blood in the placental villi to adjacent arterial blood), local inequalities between maternal and fetal placental blood flow[13,14] (analogous to ventilation/perfusion inequalities in the lung), and a high rate of oxygen consumption by the placenta.[15]

However, the predictions related to the orientation of maternal and fetal placental blood flows are borne out. In those species with counter-current exchangers (horse and guinea pig) $Po_2$ in the umbilical vein and artery are higher than in the other species. However, the main finding of these studies is that while the fetus receives adequate amounts of oxygen to meets in growth and metabolic needs, the $Po_2$ in arterial blood is much lower than in the adult. A negative consequence of this is that the oxygen gradient between blood and cytochrome C on the inner mitochondrial membrane is lower in the fetus than in the adult, which means that the fetus is much lower to the minimal gradient below which oxygen diffusion from blood to tissue is impaired. However, there appears to be some reserve in the placental transport of $O_2$, at least in the fetal lamb in late gestation. Intravenous infusion of norepinephrine for 50 min[16] or triiodothyronine for 5 days[17] results in a ~25% increase in fetal $O_2$ consumption, associated with significant increases in umbilical blood flow and placental diffusing capacity of $O_2$ (with norepinephrine administration) and no significant changes in fetal blood gas or acid-base status. This reserve can accommodate the transient increases in fetal $O_2$ consumption with episodic fetal activity[18,19]. Moreover because uterine blood flow does not change with transient changes in fetal $O_2$ consumption, an increase in the latter is associated with a fall in vascular $Po_2$[18-20], whereas a decrease is associated with a rise in $Po_2$[21-23]. The transient falls in vascular $Po_2$ of course lower the gradient for oxygen diffusion.

1.4  Compensatory Mechanisms for the Low Vascular Po$_2$

   1.4.1 Higher O$_2$ Affinity of Fetal Blood

There appear to be at least two mechanisms that in part compensate for the low Po$_2$ of fetal blood. The best studied is a higher O$_2$ affinity of fetal hemoglobin or blood compared to the mother (Figure 1). There are 2 features of the fetal curve that are of benefit to the fetus. The first is the higher affinity for oxygen, which means that any Po$_2$ value oxygen saturation is higher in the fetus. The second feature is where on the curve the fetus operates. This is on the steep portion, in contrast to the adult, so that small changes in Po$_2$ result in larger changes in oxygen saturation.

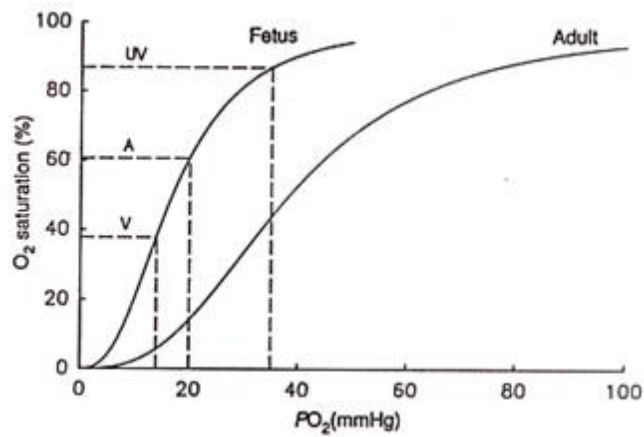

**Figure 1. O$_2$ dissociation curves for fetal and maternal blood in sheep. The line intersecting the fetal curve denote Po$_2$/O$_2$ saturation relationships for umbilical venous (UV), umbilical arterial (A) and systemic venous (V) bloods. (Rurak, unpublished data)**

While the above features of the fetal hemoglobin-oxygen dissociation curve seem to be of obvious survival benefit, it seems that they are not essential for fetal survival at least in some species. This is certainly the case in the human as is shown by the results of antenatal treatment of hemolytic disease of the fetus and newborn, which occurs in pregnancies with an Rh-negative mother carrying an Rh-positive fetus[24]. In some cases, this can result in severe fetal anemia and an important antenatal treatment is intrauterine transfusion of adult red cells. In severe cases this can result in near complete replacement of fetal with adult red cells. Perinatal survival for this treatment averages 90% and follow up studies have found normal development in more than 95% of the cases, whereas without the treatment many of the affected fetuses would have died *in utero*. Even in sheep, which have the largest difference between fetal and adult hemoglobin oxygen affinity[25], at least some fetus can survive near complete replacement of fetal blood with adult blood[26,27].

### 1.4.2 High Fetal Combined Ventricular Output

The other compensatory mechanism is a much higher resting cardiac output in the fetus compared to the adult. In the human, the fetal combined ventricular output has been measured by Doppler ultrasound measurements of blood flows through the tricuspid and mitral valves in the heart[28], whereas in the sheep it has been measured with radioactive[29] or fluorescent microspheres[30] or by implanting flowmeters on the outflow vessels of the heart[31,32]. In both species, combined ventricular output in the fetus averages ~450 ml/min.kg body weight[31], which is much higher than the value of ~100 ml/min.kg in the adult[33,34]. Even when the portion of output that supplies the placenta and not fetal tissues is subtracted, the remaining "systemic flow: is 337 ml/min.kg in the human and 250 ml/min.kg in the sheep. The higher value is the human is due to a lower umbilical blood flow, but as vascular $Po_2$ and $O_2$ content are higher in the human, oxygen delivery to the fetus is similar in the two species. When results from other species are examined, this relationship between umbilical blood and vascular $Po_2$ is maintained as shown in Figure 2.

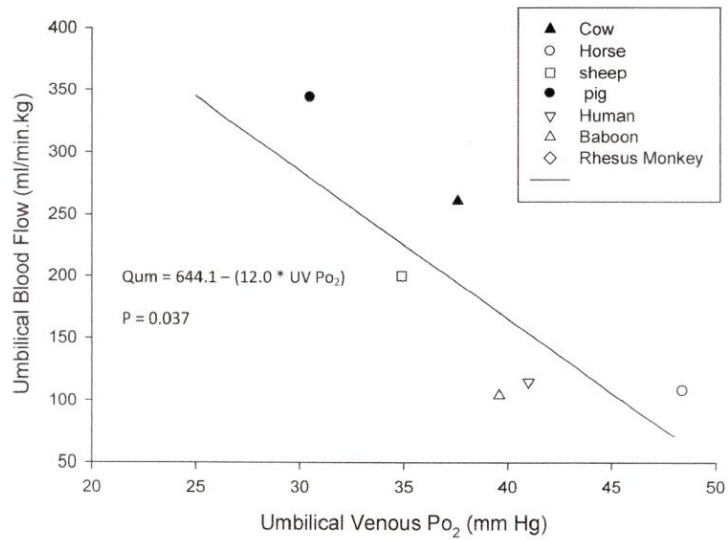

Figure 2. The relationship between umbilical venous $Po_2$ and umbilical blood flow in several species. Data are from the following sources: cow: Reynolds et al, 1985; Comline and Silver[8], horse: Fowden et al[88], Comline & Silver[8]; sheep: Rurak unpublished data; pig: Harris et al[229]; Comline and Silver[8]; human: Sutton et al[73], Soothill et al[80]; baboon: Fisher et al[230]; Rhesus monkey: Lo et al[232].

The high cardiac output results in higher rates of blood flow to nearly all fetal tissues and organs. This reduces the decrease in blood $Po_2$ as it passes through tissues. The high output results from both a high heart rate and stroke volume[31]. The downside of a high resting cardiac output is that the fetus, at least the fetal lamb, has very limited ability to increase output[35]. There is limited information on fetal cardiac output from other species. However, the fetal heart rate in other species is similar to the sheep and human, and the inverse relationship between heart rate and body weight observed in adults is not observed in the fetus[36].

1. 5 Negative Consequence of a Low Vascular $Po_2$

The results discussed above indicate that the fetus receives adequate amounts of oxygen but at a low $Po_2$ and that the fetus is at risk of further reductions in vascular $Po_2$, which could limit diffusion of oxygen from blood to tissue. Towell[37] measured tissue $Po_2$ in fetal lambs with an implanted galvanic electrode. Several tissues were examined including the dura, subcutaneous tissue and muscle. During induced fetal hypoxemia (achieved by lowering maternal inspired $O_2$ to 10%) tissue $Po_2$ in subcutaneous tissue felt from ~8 mm Hg to ~4-5 mm Hg and in muscle from ~4.5 mm Hg to ~2.5 mm Hg. Koos and Power[38] calculated brain tissue $Po_2$ in fetal lambs using equations based upon the Krogh cylinder model of oxygen diffusion. The brain $Po_2$ was estimated to be 12 mm Hg. Hypoxemia decreased the predicted $Po_2$ to 7.6 mm Hg, while anemia decreased it to 8.7 mm Hg. Gu et al[42] conducted studies on chronically instrumented fetal lambs in which cerebral oxygen consumption was measured with variations in vascular $Po_2$. They found that cerebral oxygen consumption was not correlated with sagittal vein $Po_2$ over the ranges of 9-25 mm Hg. A sagittal vein $Po_2$ of 9-11 mm Hg corresponds to an arterial $Po_2$ of 14 mm Hg. The sagittal vein $Po_2$ of 9 is higher than the predicted brain $Po_2$ of 7.6 mm Hg during hypoxemia[38,39], although this latter value has to be verified experimentally. Nonetheless, the results suggest that reductions in fetal oxygen delivery that also result in fetal hypoxemia would be less well tolerated than reductions that do not. Rurak et al[39] studied the effects of prolonged fetal hypoxemia (achieved by lowering maternal inspired oxygen to 9-10%) whereas Kwan et al[40] studied the effects of acute 40-45% fetal hemorrhage. Some of the results of these 2 studies are shown in Figure 3.

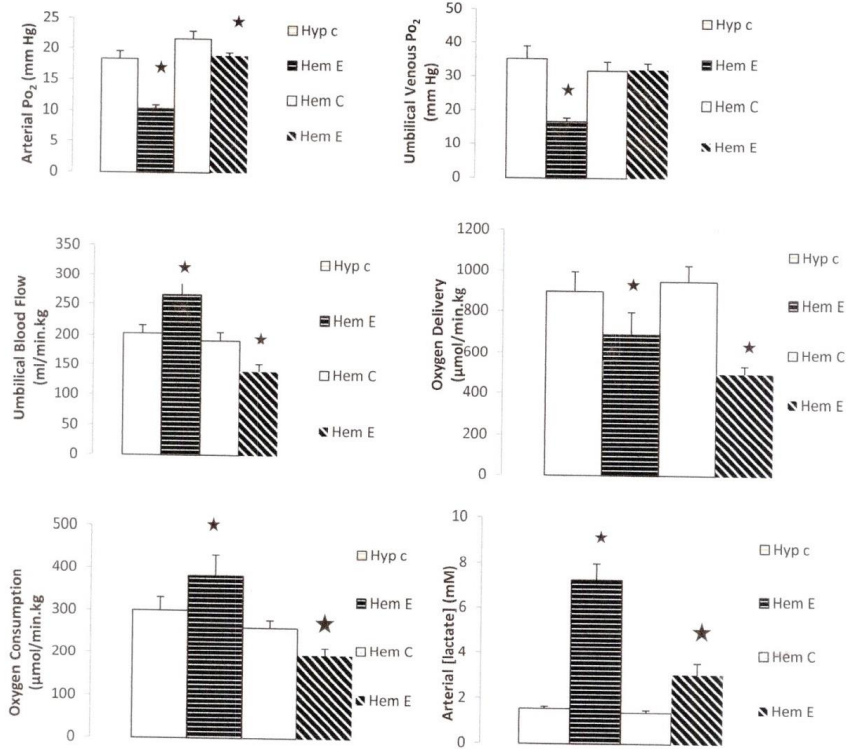

Figure 3. Plots of control and experimental for arterial Po$_2$, umbilical venous Po$_2$, umbilical blood flow, oxygen delivery, oxygen consumption and arterial lactate concentration after 1.7 h of feta; hypoxemia and 2 h after 40-45% acute hemorrhage. Hyp C = hypoxia control; Hyp E = hypoxia experimental; Hem C = hemorrhage control; Hem E = hemorrhage experimental. From Rurak et al[39]; Kwan et al[40].

The values were obtained 1.7 h into the hypoxic period and 2 h after the end of hemorrhage. The fall in arterial Po$_2$ was greater with hypoxemia (18.4±1.2 to 10.4±0.5 mm Hg) than with hemorrhage (21.7±1.3 to 19.2±0.5 mm Hg). Umbilical venous Po$_2$ also fell during hypoxia whereas there was no change with hemorrhage. Umbilical blood flow rose during hypoxia, perhaps due to a rise in cardiac output due to an increase in circulating catecholamine concentrations[39]. In contrast, umbilical blood flow decreased following hemorrhage, due to a fall in cardiac output[40]. Fetal O$_2$ delivery fell during both perturbations but the fall following hemorrhage (48%) was greater than with hypoxemia (23%). There were opposite effects on fetal $O_2$ consumption, with an increase with hypoxemia and a decrease following hemorrhage. The fall during hemorrhage is likely the result of the large fall in $O_2$ delivery. An explanation for the rise during hypoxemia is not obvious, but may be due to rise in oxygen radicle production, which increase oxygen consumption without increasing ATP concentration. Arterial lactate concentration increased in both experiments, with the larger rise occurring during hypoxemia. This is somewhat surprising since at this time during the hypoxemia oxygen consumption had increased, whereas at 2 h after hemorrhage oxygen consumption had declined meaning that some fetal tissues were receiving inadequate delivery of oxygen and presumably had to switch to anaerobic metabolism leading to increased lactate production. Other researchers have reported increased blood lactate levels with fetal hypoxemia[41-43]. Gu et al[43] also reported that the lactate/pyruvate ratio rose during fetal hypoxemia achieved by reducing uterine blood flow for 60 minutes, indicating that the rise in lactate was due to tissue hypoxia. They also found that fetal oxygen consumption did not fall until umbilical venous $Po_2$ fell below 15 mm Hg. Milley[44] subjected pregnant ewes to 3 h of hypoxemia, achieved by lowering the inspired $O_2$ concentration and found that the normal uptake of fetal lactate from the placenta[15] ceased, indicating the increase in blood lactate concentration came from fetal tissues. Rurak et al[39] suggested that the rise in lactate concentration initially during hypoxemia resulted from a rise in circulating catecholamines since increased lactate was not associated with hypoxia. In addition, epinephrine infusion to fetal lambs results in an increase in blood lactate concentration[45] and the rise in lactate during hypoxemia can be partially blocked by propranolol[42]. However, fetal hemorrhage also results in a rise in noradrenaline concentration and as shown in Figure 3, the rise in arterial lactate concentration was much less following hemorrhage than with hypoxemia. Moreover at 2 h following hemorrhage fetal $O_2$ consumption had decreased, that is there was hypoxia and yet the rise in lactate was modest. During the hypoxemia experiments the blood lactate concentration rose progressively and at the end of the experiment at 7.1 h arterial lactate had risen to 25.6 mM with an arterial pH of 6.483. The rise in lactate decreased hemoglobin $O_2$ saturation via the Bohr effect, which led to a progressive decrease in blood $O_2$ content. At 7.1 h, arterial $O_2$ content had decreased to 0.4 mM, an 86% fall from the control value. Umbilical blood flow, fetal $O_2$ delivery and fetal $O_2$ consumption fell progressively following the 1.7 h samples. At 7.1 h $O_2$ consumption had decreased significantly from the control value and $O_2$ delivery had fallen by 73%. In this study the animals were sacrificed after the last sample so that long term outcome could not be determined. However, two of the fetuses died shortly after the last sample. In contrast, on the first day following hemorrhage, fetal $O_2$ consumption and arterial lactate concentration were not different from the control values. All fetuses survived the hemorrhage and were born alive. However, fetal growth ceased following the hemorrhage[46].

The results discussed above indicate that the fetal lamb cannot survive severe hypoxemia, compared to its ability to deal with acute hemorrhage. The precipitating event of the hypoxemia study was the fall in vascular $Po_2$. As mentioned above arterial $Po_2$ had fallen to 10.4 mm Hg at 1.7 h of hypoxemia. That fall in $Po_2$ would also result in a large decrease in blood oxygen saturation and thus blood $O_2$ content, because of where the fetus operates in the hemoglobin-oxygen dissociation curve (Figure 1). In our experience, fetal lambs which spontaneously and chronically exhibit an arterial $Po_2$ of ~10 mm Hg, do not survive until term. It seems that a $Po_2$ of 11 or greater is necessary for fetal survival (Figure 4). This also seems to be the case for growth restricted human fetuses at least as based upon cordocentesis

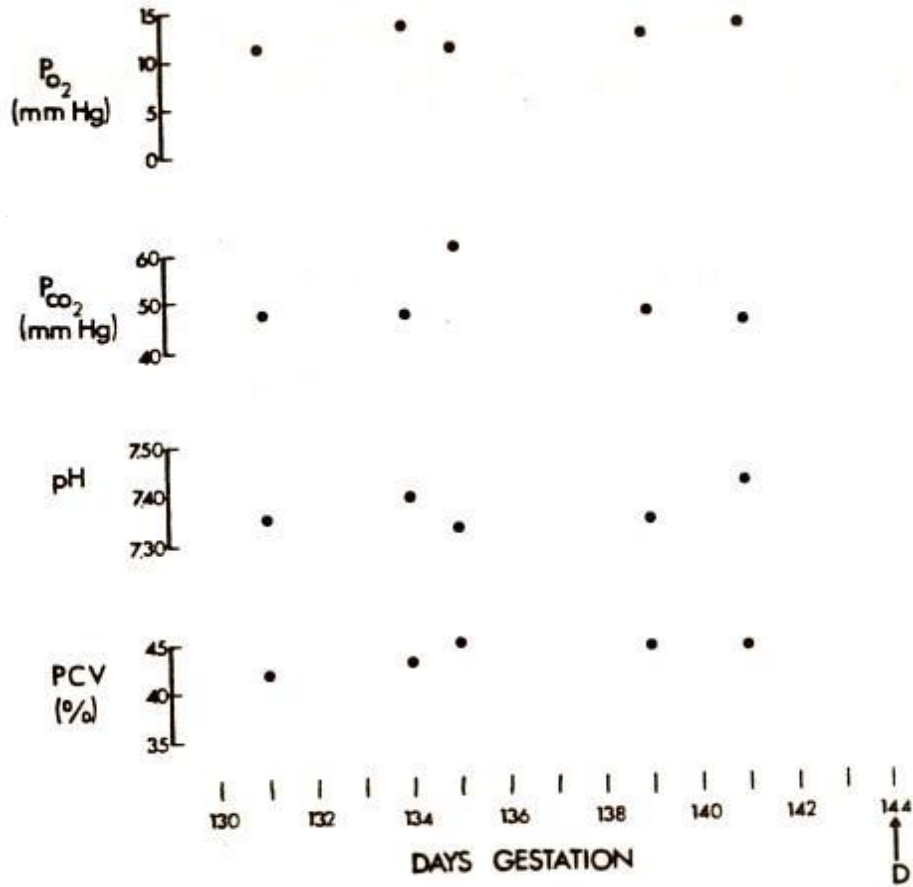

Figure 4. Arterial $P_{O_2}$, $P_{CO_2}$, pH and hematocrit (PCV) in a chronically instrumented, spontaneously growth-restricted fetal lamb in late gestation. (Rurak, Unpublished data)

samples from such fetuses[47]. Whether the human fetus has a tolerance to hemorrhage similar to the fetal lamb is uncertain. Blood loss from the human fetus occurs most commonly from fetomaternal hemorrhage, which is the entry of fetal blood into the maternal circulation[48]. While this can be detected by identifying fetal red cells in maternal blood, determining the volume of fetal blood lost requires formulae that include maternal and fetal hematocrit and maternal blood volume and the latter two have to be assumed. Moreover, no information is provided on the timing and rate of hemorrhage and the latter value is of critical importance in determining fetal survival[48].

To summarize, the fetus receives adequate amounts of oxygen from the mother, but vascular $Po_2$ is lower, which lowers the gradient for oxygen diffusion from plasma to the inner mitochondrial membrane, and results in the fetus being closer to the minimal gradient for diffusion of $O_2$ from blood to tissue. It also seems that during severe fetal hypoxemia vascular $Po_2$ may fall low enough to impair diffusion.

2.0 Changes in Fetal Oxygenation during Gestation

Our interest in this subject was prompted by epidemiologic studies of the rate of human stillbirth, fetal growth restriction and other perinatal complications as a function of gestational age. When stillbirths at a given gestational age are divided by all fetuses at risk of stillbirth at that gestational age, including those that deliver later in gestation (fetuses-at-risk approach), there is a gestational age-related increase in the incidence of stillbirth and several other adverse pregnancy outcomes[49]. However, these findings are not revealed when the stillbirth rate is expressed as the number of stillbirths at a given gestational age divided by all births at that gestational age (conventional approach).

In fact, an opposite finding is obtained using this conventional approach: the risk of stillbirth declines with advancing gestational age[50]. The fetuses-at-risk approach was proposed in 1987 by Yudkin et al[51]. More recently, this formulation has been extended to other pregnancy-related phenomena and shows that rates of pregnancy complications, fetal growth restriction and perinatal death increase at late gestation[52] (Figure 5).

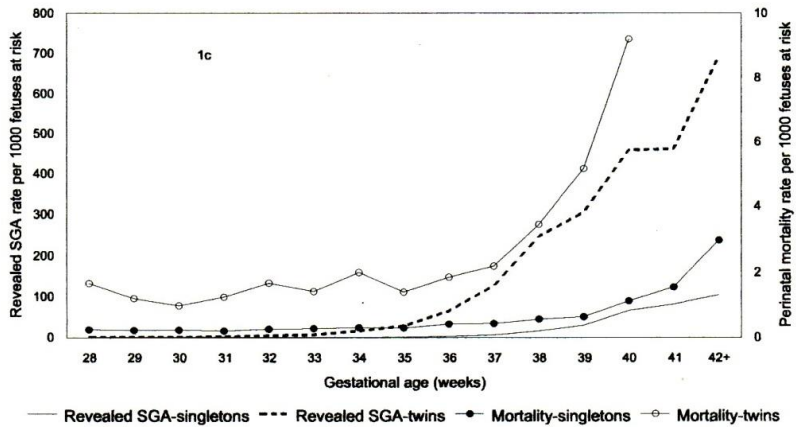

Figure 5. The relationship between gestational age and the rate of stillbirth and small for gestational age in human singleton and twin pregnancies. From Joseph[49].

This proposition is supported by evidence from randomized clinical trials; the DIGITAT trial showed that labour induction leads to earlier delivery and substantially lower rates of intrauterine growth restriction[53], while meta-analyses of randomized clinical trials on labour induction for post-term pregnancy show that such early delivery reduces perinatal mortality compared with expectant management[54]

The relationship between gestational age and fetal growth restriction and stillbirth, and the association between fetal growth restriction and stillbirth, suggest a causal mechanism: that with advancing gestational age, the ability of the maternal-placental-fetal nutrient supply line becomes increasingly unable to meet fetal metabolic requirements due to the rapidly increasing size of the fetus in late gestation. This is not a new idea. Hippocrates of Chios in the 5$^{th}$ century BC stated in his treatise on the Nature of the Child: "My assertion is that what brings on birth is a failure in food supply"[55]. More recently, Barcroft et al[56] reported studies in anesthetized pregnant ewes, in which fetal arterial $O_2$ saturation was measured from 60 days gestation to term (~147 d). The value was low in early pregnancy, rose until ~120 day and declined progressively thereafter. This led Barcroft to conclude:

"…from the 120$^{th}$ day onwards…. the conditions deteriorate. The fetus is approaching a crisis. That crisis is the alternative between death and escape."[57] However, the subsequent development of techniques for the chronic instrumentation of pregnant sheep, which allowed the study of the undisturbed fetus *in utero*[58] raised questions about the extent to which the results obtained from acute preparations, such those as studied by Barcroft et al[56], were affected by their acute nature, particularly anesthesia, and these results have been largely ignored by subsequent researchers.

As discussed earlier, fetal vascular $P_{O_2}$ is much lower than after birth and cannot fall much lower without affecting diffusion of $O_2$ from blood to tissue. Also, oxygen is continuously needed for fetal survival, there are very limited oxygen stores in the body, there are no alternatives to oxygen, in contrast to other molecules required by the fetus (e.g. glucose) and maternal to fetal oxygen transfer depends primarily on the magnitudes and relative orientations of maternal and fetal placental blood flows[5,11]. In addition, the previously mentioned studies of Barcroft and Kennedy suggest that at least in sheep fetal oxygen supply could become limiting in late gestation[56,57]. Recently Pacaro et al measured the concentrations of erythropoietin (a measure of fetal hypoxia), cardiac troponin I (a measure of myocardial damage) and glial fibrillary acidic protein (a measure of brain damage) in amniotic fluid samples collected via amniocentesis in 60 cases associated with stillbirth of a structurally normal fetus and in 60 controls[59]. Fetal hypoxia was present in 88% of the stillbirths and in 91% of these cases, heart and/or brain damage was also present. There are also the reports of fetal hypoxemia associated with fetal growth restriction in humans[47] and in animal models of fetal growth restriction[60,61].

2.1 Variables that Determine Maternal-Fetal Oxygen Transfer and Fetal $O_2$ consumption

Thus, the variables that determine maternal to fetal oxygen transfer and fetal oxygen consumption are of primary interest. On the maternal side, these include maternal arterial oxygenation and uterine blood flow. However, although maternal arterial $P_{CO_2}$ decreases in pregnancy and this leads to an increase in pH, the impact on arterial $P_{O_2}$ and $O_2$ saturation appears to be minimal[62]. Thus, maternal oxygenation is likely not a limiting factor unless there are maternal cardio-respiratory problems or with pregnancies at high altitude[63]. On the fetal side, the relevant variables include umbilical blood flow and fetal blood oxygenation. Fetal oxygen delivery can be estimated as the product of umbilical blood flow and umbilical venous oxygen content, whereas fetal oxygen consumption is the product of umbilical blood flow and the umbilical veno-arterial difference in $O_2$ content. Fetal oxygen consumption can also be estimated by measuring maternal $O_2$ consumption before and after delivery and taking the difference between these estimates[64]. In terms of the contributors to fetal $O_2$ requirements, two factors which are potentially modifiable, include fetal breathing and body movements, and fetal growth. For all these variables, longitudinal measurements for a least of portion of gestation are important.

2.2 Uterine Blood Flow

Uterine blood volume flow in human pregnancy has been measured using Doppler ultrasound. Total flow increases progressively during gestation[65]; however, when uterine blood flow is normalized to estimated fetal *in utero* weight, it declines progressively, at least from 20 weeks gestation (Figure 6)[66].

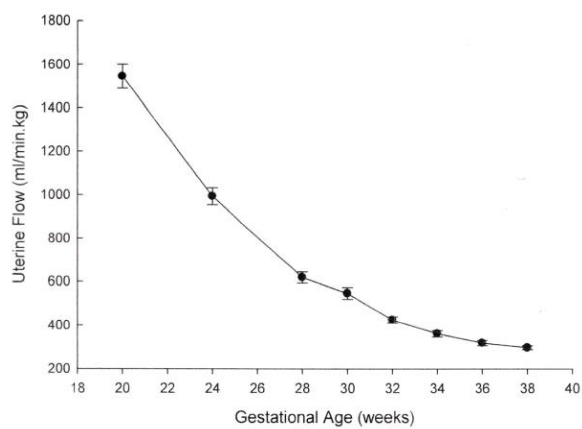

Figure 6. Uterine blood flow normalized to fetal weight (mean±SE) as a function of gestational age in human pregnancy. Data plotted from Konje et al[66].

In pregnant animals, uterine blood flow has been measured longitudinally in a number of species including sheep[67-69], horses[8], rabbits[70] and guinea pigs[71]. In all species, total uterine blood flow increases progressively during gestation. However, when normalized to fetal weight, uterine blood flow falls progressively, as in the human (Figure 7). And in pregnant sheep whose gestation was prolonged by fetal hypophysectomy, weight-normalized uterine flow continued to decrease (Figure 7).

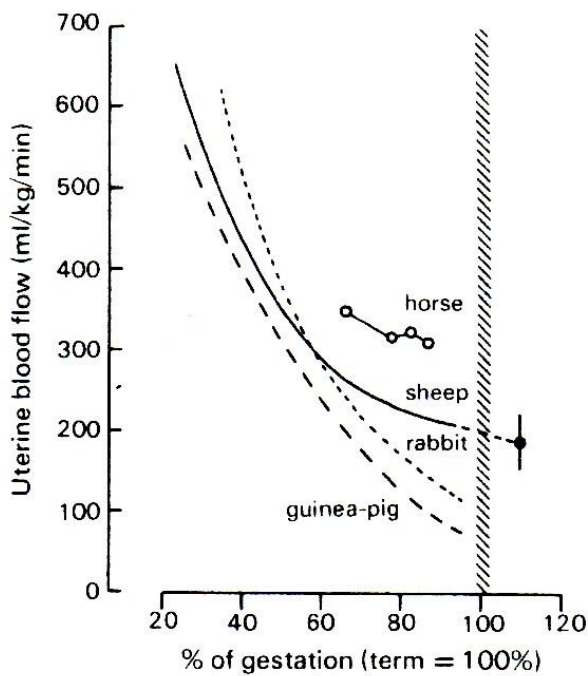

Figure 7. Uterine blood flow as a function of gestation normalized to fetal weight for several mammalian species. Reproduced by the kind permission of Recent Advances in Physiology, from Silver et al 1982.

2.3 Umbilical Blood Flow

In human pregnancy, umbilical blood flow has been measured longitudinally in several studies using Doppler ultrasound[72-74]. All these studies report a decrease in umbilical blood flow normalized to fetal weight (Figure 8), although Sutton et al[73] found that the change was not statistically significant. Gill et al[72] reported that total umbilical blood flow increases progressively until about 38 weeks gestation and then plateaus or perhaps decreases slightly. When related to fetal weight, flow begins to decline at ~36 weeks gestation. Data on the longitudinal changes in umbilical blood flow in animal species (sheep, cow, foal) are summarized in Silver et al[75]. In all the species, umbilical blood flow/kg fetal weight decreased with advancing gestational age, and, in pregnant sheep in which gestation length was prolonged by fetal hypophysectomy or adrenalectomy, flow decreased further. In Figure 9 values of umbilical blood flow/kg fetal weight collected in the lab of one of us (DWR) are presented as a function of gestational age. Data from 10 chronically instrumented fetal lambs in which flow was measured at a single point in time or followed longitudinally are included. Linear regression analysis indicated a significant decrease in weight-normalized umbilical blood flow and the slope of the regression line is similar to that reported by Hedriana et al[76].

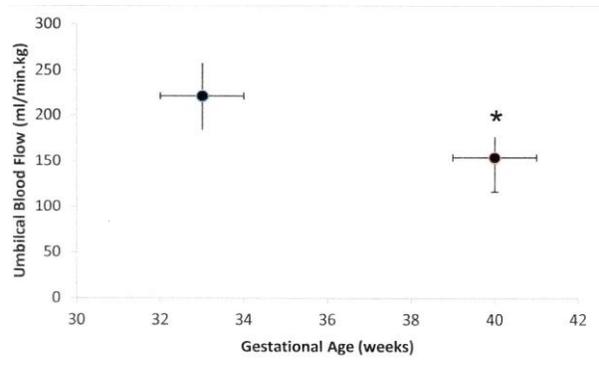

Figure 8. Human umbilical blood flow normalized to fetal weight as a function of gestational age. Plotted from Link et al, 2007

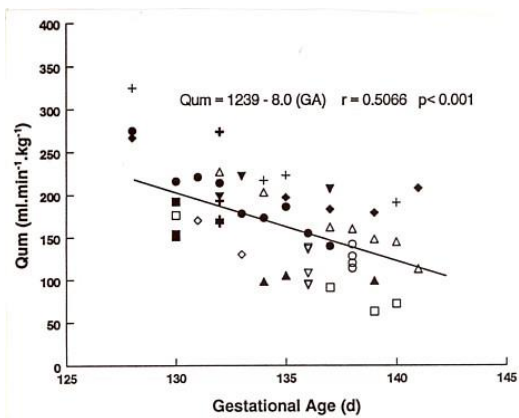

Figure 9. Umbilical blood flow/kg fetal weight in the fetal lamb as a function of gestational age. Each symbol represents a different fetus. D Rurak, unpublished data.

2.4 Fetal Blood gas and Acid-Base Status

In pregnant sheep, acute experimental reductions in uterine or umbilical blood flow decrease fetal arterial oxygenation and pH and increase $Pco_2$[77-79]. Thus, it is reasonable to expect that the changes in flow in these vessels during gestation should result in similar, progressive changes in fetal blood gas and acid-base status. Much of the available data comes from human cross-sectional studies, in which blood gas and acid-base parameters were measured in umbilical arterial and venous blood samples obtained via cordocentesis from patients over a wide gestational age range. The results indicate decreases in $Po_2$ and $O_2$ saturation and increases in $Pco_2$ and hemoglobin concentration or hematocrit with advancing gestation[80-84] (Figure 10). Soothill et al[80] reported gestational age-related decreases in $Po_2$ and pH and increase in $Pco_2$ in umbilical venous, umbilical arterial and intervillous blood. In terms of fetal blood gas and acid-base data from animal species, the results of Barcroft et al[56] have already been mentioned: in anesthetized pregnant ewes, fetal arterial $O_2$ saturation increased from 60 to ~120 days gestation and then declined progressively. However, subsequent studies on chronically instrumented fetal lambs and foals have not confirmed these findings, in that there were no changes in blood gas and acid-base variables during gestation[58,85-89]. However, the sampling approaches used in these studies differed from those in the human cordocentesis studies mentioned above, in that either individual fetuses

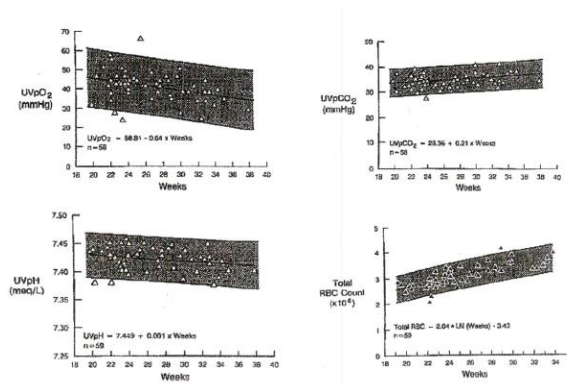

Figure 10. Values of umbilical venous $Po_2$ (UVpO2), Pco2 (UVpCO2), pH (UVpH) and total red cell count as a function of gestational age in human fetuses subject to cordocentesis. Adapted by the kind permission of Obstetrics and Gynecology, from Weiner et al[81]

sampled longitudinally over a relative short range of gestational age[85-87] or small groups of fetuses were studied at different gestational ages [58,88,89]. Given that in the fetal lamb vascular $Po_2$ and other blood gas and acid-base variables exhibit marked short term fluctuations in association with fetal somatic activity and pre-labor uterine contractions[18,90-93], it is unlikely that either of the experimental approaches utilized in the sheep and horse studies mentioned above could have detected any gestational age-related changes in fetal blood gas and acid-base status. Recently, we examined blood gas and acid-base variables in 447 control, arterial blood samples collected from 108 chronically instrumented fetal lambs between 103 and 146 days gestation[94]. With advancing gestation, $Po_2$, pH, $O_2$ saturation and $O_2$ content fell significantly, while $Pco_2$ and hemoglobin concentration increased (Figure 11).

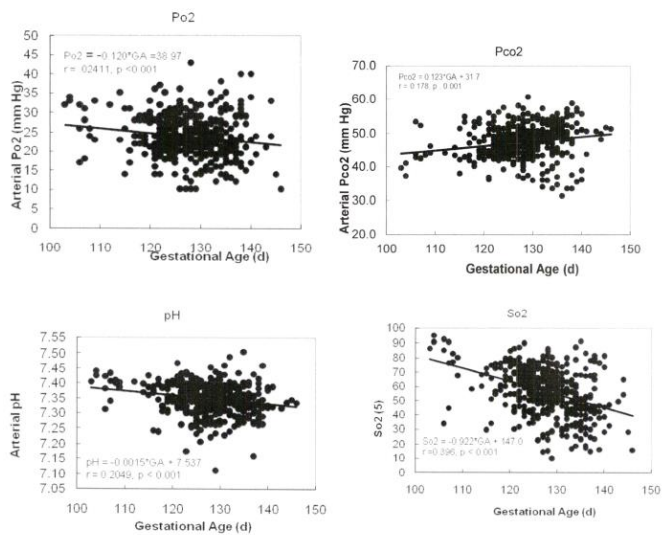

Figure 11. Values of fetal arterial Po2, Pco2, pH and $O_2$ saturation as a function of gestational age in 447 blood samples from 108 pregnant ewes from 103 to 146 d gestation. From Rurak et al[94]

Table 2 presents the slopes of the variables plotted in the figure along with the calculated values for 100 and 147 (term) day's gestation. Arterial $Po_2$ fell by 21% over this gestational age range, whereas the fall in oxygen saturation (53%) was much greater due to the progressive rise in $Pco_2$ and fall in pH and also because the fetus operates on the steep portion of the hemoglobin-oxygen dissociation curve Figure 1)..

The decrease in blood $O_2$ content (37%) was less due to the progressive increase in hemoglobin concentration. Thus, both in the human and sheep, fetal oxygenation decreases with advancing gestation.

Table 2. Slopes of the regression equations calculated for the relationships between gestational age and arterial $Po_2$, $Pco_2$, pH, base excess (BE), % hemoglobin oxygen saturation ($So_2$), hemoglobin concentration (Hb) and blood oxygen content ($Co_2$) in fetal lambs. Also presented are the values calculated for 100 and 147 d gestation

| Variable | Slope ±SE (p) | 100 d | 147 d |
|---|---|---|---|
| $Po_2$ (mm Hg) | -0.119±0.036 (<0.001) | 27.1 | 21.5 |
| $Pco_2$ (mm Hg) | 0.123±0.032 (<0.001) | 44.0 | 49.5 |
| pH | -0.155±0.0003) (<0.001) | 7.390 | 7.321 |
| BE (meq/L) | -0.054±0.019 (<0.001) | 2.6 | 0.0 |
| $So_2$ (%) | -0.922±0.012 (<0.001) | 81.8 | 38.5 |
| Hb (g%) | 0.065±0.011 (<0.001) | 8.7 | 11.7 |
| $Co_2$ (mM) | -0.037±0.007 (<0.001) | 4.6 | 2.9 |

5 Fetal Oxygen Delivery and Consumption

As mentioned previously, fetal oxygen delivery can be estimated as the product of umbilical blood flow and umbilical venous $O_2$ content. In the human, fetal oxygen delivery has been estimated in preterm and term infants by measuring umbilical blood flow with pulsed Doppler ultrasound 1 to 126 h (median 6-8 h) prior to delivery and multiplying this by the $O_2$ content measured in a cord venous blood sample collected immediately after delivery[95]. Fetal $O_2$ delivery at 33 and 40 weeks gestation averaged 26.6±9.2 and 17.6±6.7 (ml/min)/kg, respectively. Measurements of fetal $O_2$ delivery and consumption have been obtained from a number of animal species, particularly from the sheep[69,96] and horse[88]. $Do_2$ (oxygen delivery) and $Vo_2$ (oxygen consumption) values from fetal lambs at different gestational ages have been reported in several studies[18,39,96-99]. Both variables decrease progressively and significantly with advancing gestational age. However, longitudinal measurements of $Vo_2$ as a function of gestational age have reported only by Bell et al[98], Fowden et al[88] and Molina et al[99] (Figure 12).

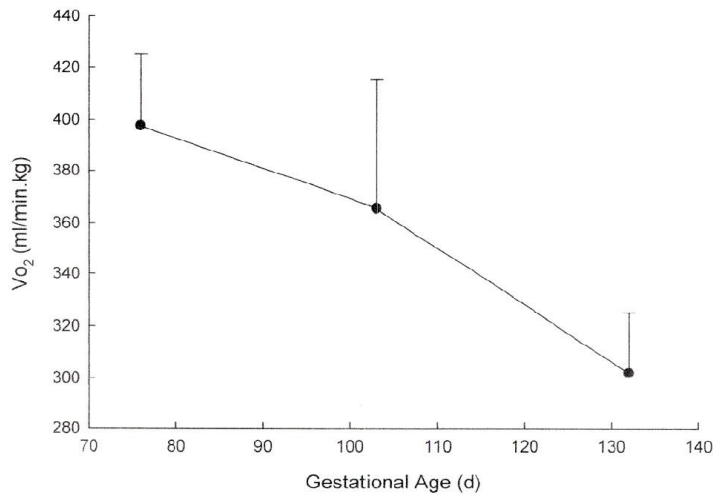

Figure 12. Mean±SE values of $O_2$ consumption in fetal lambs in relation to gestational age. Plotted from Molina et al[99].

In fetal lambs at mid-gestation (73-97 d), $Vo_2$ was significantly higher than the values obtained later in gestation (119-141 d) and the % decrease was 24.1. In the fetal horse, longitudinal measurements at 5 gestational ages between ~ 220 and 320 d are available. $Vo_2$ falls more or less progressively during gestation[88].

2.6 Nature of the Fetal $O_2$ Demands during Gestation

From the discussion above, it is clear that in the species, for which data are available, fetal $O_2$ delivery falls progressively during gestation and in the sheep and horse this is associated with a fall in fetal $Vo_2$. Except for relatively short periods of time, overall $O_2$ delivery and consumption must be in balance, otherwise oxygen debt and tissue hypoxia will develop. Thus, as gestation proceeds, fetal $O_2$ requirements must be reduced at the same rate as $O_2$ delivery falls. In the fetus, the main uses of oxygen are for maintenance functions (e.g. fetal and placental cellular transport functions, cardiovascular and renal functions, neuronal activity, biosynthetic activities), fetal breathing and body movements and growth, and for the human fetus, this latter process includes the deposition of significant amounts of adipose tissue in late gestation[100]. Fetal breathing movements are important for lung growth and maturation[101], while fetal body movements are important for normal development of the musculo-skeletal system[102]. Together these latter activities account for ~8-15% of total fetal $O_2$ consumption[18,21-23].

2.6.1 Fetal Breathing and Body Movements

Breathing and body movements and behavioral states in the human fetus have been documented in many studies using real time ultrasound observation[103-106]. Longitudinal measurements of fetal body movement in 29 singleton pregnancies from 24 to 40 weeks gestation have been reported by Ten Hof et al[107]. As illustrated in Figure 13, although there is considerable inter-subject variation in the overall amount of movement, the percentage of fetal movements and fetal movements/h decreased progressively and the onset-onset interval (i.e. fetal quiescence) increased progressively with advancing Our group published a longitudinal ultrasound study of motility in fetal lambs that were monitored for 30 or 60 min at weekly intervals from 55 d gestation to term[108]. Figure 14A illustrates the average number of body movements/min as a function of gestational age. The movement rate was relatively constant from 55 to ~95 d gestation, and then declined progressively. As the body movement data for all fetuses studied conformed to this pattern, piece-wise regression with 2 elements was used to analyze the results. This analysis gives the slopes of the 2 portions of the regression curve, as well as the breakpoint, the gestational age at which the change in slope occurred. The breakpoint ranged from 64.9 to 109.6 d, with a mean value of 91.9±5.2 d. The decrease in the movement rate after the breakpoint was due to two

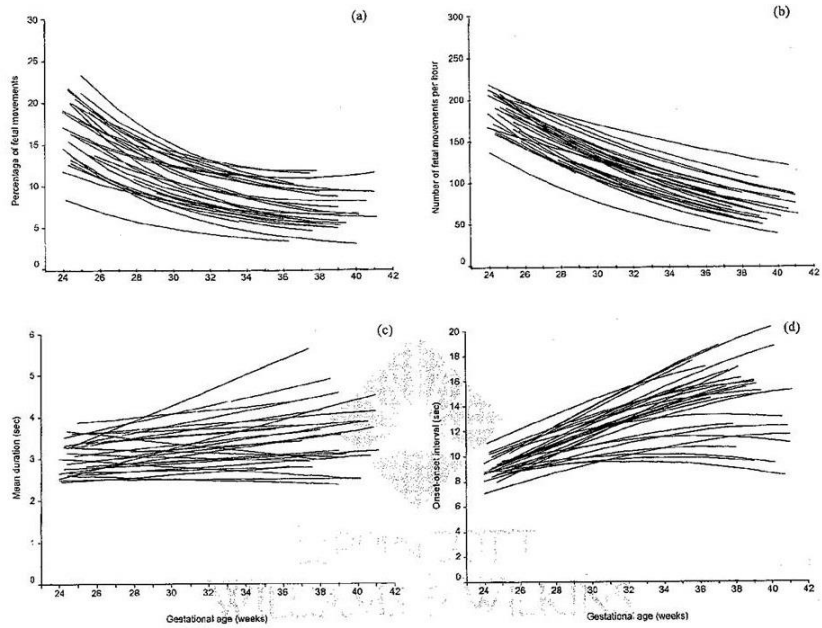

Figure 13. Regression curves for the relationship between gestational age and the percentage of fetal movements (a), the number of fetal movements (b), the mean duration of movement bouts (c) and the onset to onset interval (d) in 29 human fetuses from normal pregnancies. Reproduced by the kind permission of Pediatric Research, from Ten Hof et al[107].

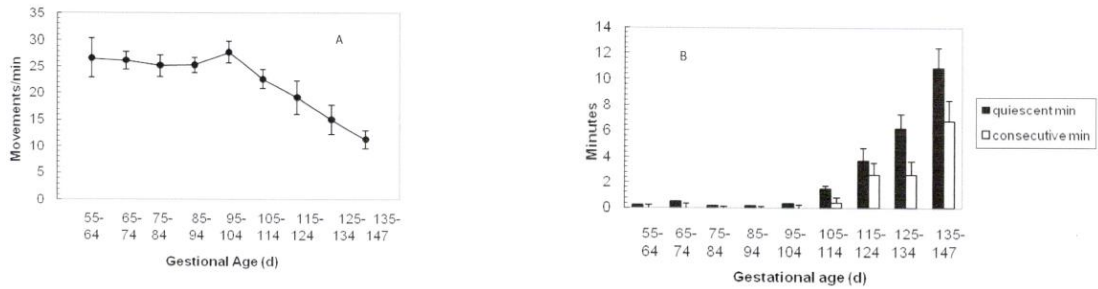

Figure 14. Mean±SE values for fetal body movements/min (A) and minutes of quiescence (B) in 8 fetal lambs as a function of gestational age. From Rurak & Wittman[108].

2 factors: a decline in the maximum movement rate from ~25/min prior to the breakpoint to ~11/min at 135-147 d, and as illustrated in Figure 14B an increase in the total number of minutes and consecutive minutes with no fetal movements (i.e. quiescence). Fetal breathing movements in the human fetus have been assessed as a function of gestational age using real-time ultrasound. There is a progressive decrease in the % time spent breathing that begins at around 34 weeks gestation[105,109]. In the fetal lamb, fetal breathing movements have been monitored longitudinally by measurement of diaphragmatic electromyographic activity[110,111] and with real-time ultrasound[108]. Both approaches have documented that breathing movements in the fetal lamb begin to decrease at around 130 d gestation (~2 weeks prior to term. Thus, in both the human and sheep, fetal motility decreases and quiescence increases in late gestation and this should decrease overall fetal $O_2$ requirements.

### 2.6.2 Fetal growth

The development of real-time ultrasound for use in human pregnancy and thus the ability to obtain 2 dimensional views of the fetus allowed for the measurement of the dimensions of various fetal structures (e.g. head, abdomen, femur, as well as crown-rump length) as a function of gestational age and use of these measurements to estimate fetal weight throughout gestation[112]. Thus, there are more data on human fetal growth than are available for other species. The studies conducted involve both cross-sectional studies with one measurement from individual subjects[113,114] and longitudinal studies with several measurements from each fetus[115-118]. A number of studies have modeled the fetal growth curve from the data obtained [113-115,119-127]. Many have examined the growth of individual structures, such as the head, abdomen and femur[113,115,121,124,125], while some have looked for changes in the trajectory of the increase in estimated fetal weight[114,122,126] or the cube values of the 2 dimensional measures of the fetal head and abdomen[120]. A consistent finding has been a reduction in growth velocity with advancing gestation. For the individual structures, the peak growth velocities occurred at ~16-17 weeks for the head diameter or circumference, 20 weeks for femur length and 20-22 weeks for abdominal circumference, with a slowing of growth after these flex points[113,124]. However, Bertino et al[124] also examined the cubic parameter of biparietal diameter, which is more proportional to head weight and found that its growth velocity peaked at 31.1 weeks (Figure 15).

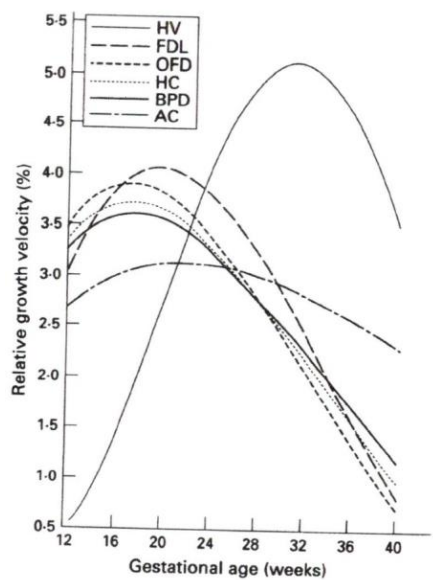

Figure 15. Mean growth velocities for head volume (HV), femur diaphase length (FDL), occipitalfrontal diameter (OFD), head circumference (HC), biparietal diameter (BPD) and head circumference (HC) (as a percentage of the size at 40 week's gestation for the human fetus. From Bertino et al[124].

Pineau et al[127] used a 2-phase regression model, which estimated a specific point in gestation at which the rate of fetal head growth decreased[119]. The value was 30 weeks gestation and there was a sex difference in fetal head growth, with males having higher values of head circumference at all gestational ages. In terms of fetal weight, Hooper et al[126] used a more complicated analytical method involving transformation of estimated fetal weights to Z scores followed by use of novel non-linear regression techniques for fetal weight data estimated from 13,593 ultrasound examinations in 7,888 pregnancies. They found that the fetal growth velocity increases up to 35 week's gestation and then decreases during the final weeks of pregnancy. Although this is at a later gestational age than was estimated for the flex point of 31.1 weeks for cubic biparietal diameter[124], the growth velocity for fetal weight is a composite value for the growth of the individual organs, and the flex points for abdominal diameter and femur length are later than the value for the fetal head. Fescina and Martell[128] and Rossavik and Deter[119] have examined the growth of the fetal cranial circumference in both the pre and postnatal periods. During pregnancy the growth is linear until about 30 weeks gestation and then declines. Following birth, the growth rate increases followed by a decrease and subsequently a linear rate of growth beginning at 4-6 months of age. This led them to conclude that the linear phases of growth in the pre and postnatal periods perhaps represent genetic influences where the terminal decrease in growth in late pregnancy and the early postnatal increase in growth are due primarily environmental influences.

There is less information on fetal growth in sheep and other species. Mellor and Matheson[129] used an implanted device to measure longitudinally the daily changes in crown-rump length in 8 fetal lambs from 100 days gestation to term. There was a linear relationship between crown rump length and fetal weight. In normal pregnancies, the daily increment in crown rump length was constant until 132 d gestation, at which point the rate decreased by 26.8% (Figure 16). In the longitudinal ultrasound study of fetal lamb somatic activity conducted by our group[108], we also measured fetal abdominal diameter longitudinally (Figure 17). The best fit of the data occurred with 2 segment linear piecewise regression. The mean breakpoint was 113.1±3.9 days gestation at which point, the slope of the regression line decreased significantly by 27% in the non-operated ewes (n=4). The breakpoint for abdominal diameter (113.1 d) was significantly greater than the breakpoint for fetal body movements (91.9±5.3 d). However, there was a linear relationship between the 2 breakpoints in the individual fetuses, i.e. the fetus with the earlier breakpoint of fetal body movements had an earlier breakpoint for abdominal diameter (Figure 18).

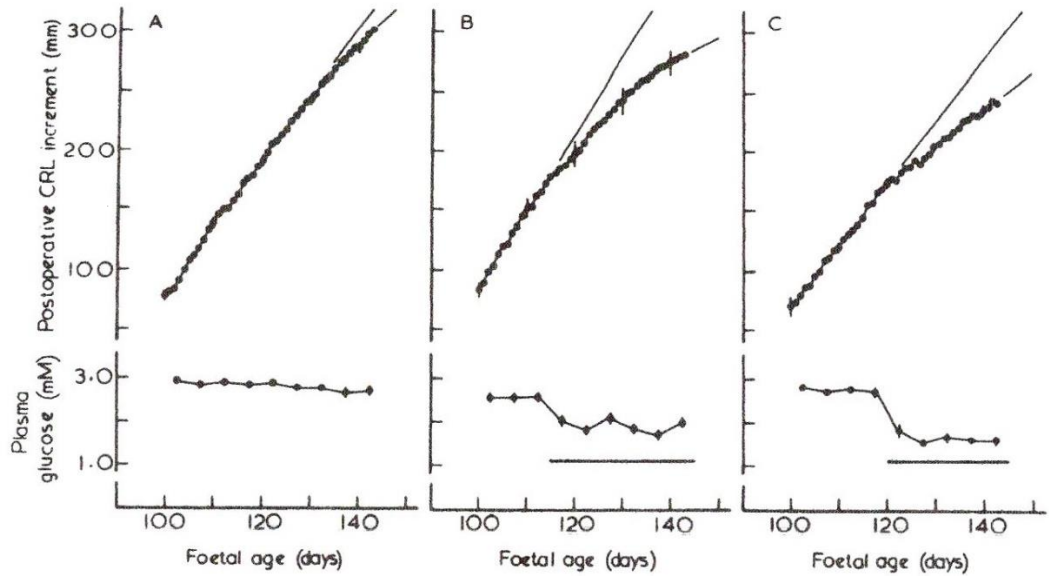

Figure 16. The mean daily crown rump length increment (upper graph) and maternal plasma glucose concentration between 100 and 143 d gestation in control pregnant sheep (A) and in moderately undernourished (B) and severely undernourished (C) animals. From Mellor et al[129]

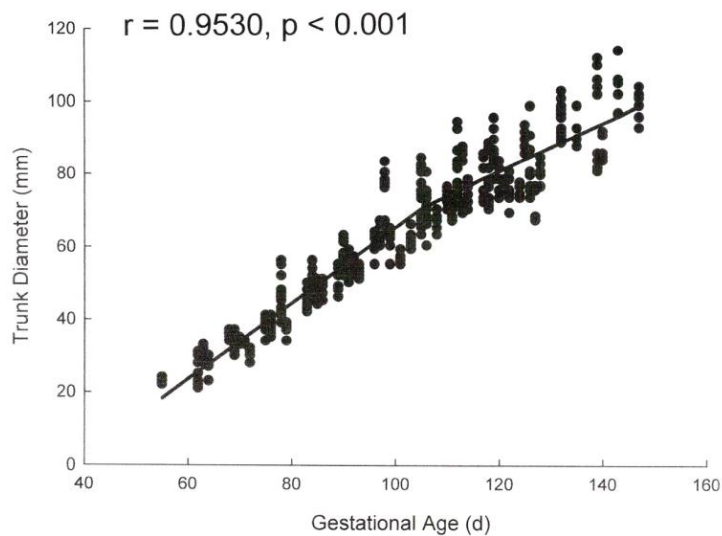

Figure 17. Longitudinal ultrasound measurements of abdominal diameter in fetal lambs as a function of gestational age. From Rurak & Wittman[108]

Thus, there are abundant data from the human to indicate that fetal growth decreases in late gestation and a lesser, but convincing amount of data from sheep to indicate the same phenomenon.

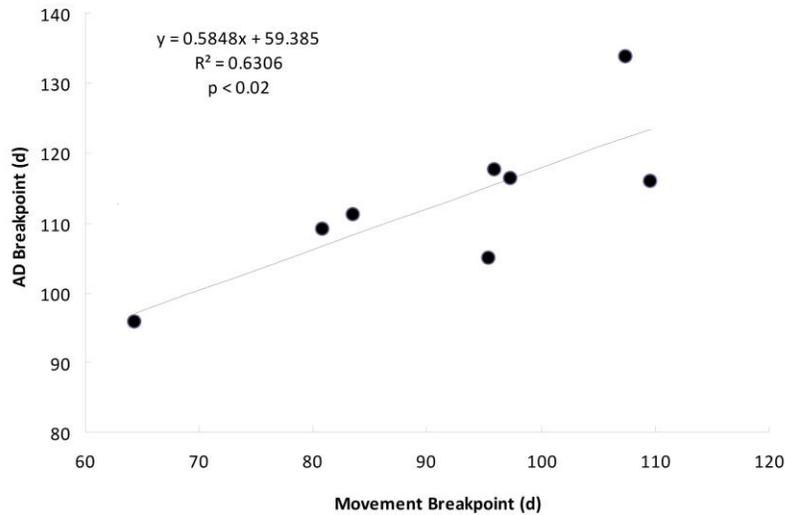

Figure 18. The relationship between the gestational age at which body movements first decreased (movement breakpoint) and the gestational age at which abdominal diameter first decreased (AD breakpoint) in 8 fetal lambs. From Rurak & Wittman[198].

2.7 Causal mechanisms

In the preceding sections, data has been presented that support the following hypothesis: in humans, sheep and several other species, uterine and umbilical blood flows normalized to fetal weight decrease with advancing gestation age. In humans and sheep, this is associated with a decrease in vascular $P_{O_2}$, $O_2$ saturation, and pH and increases in $P_{CO_2}$ and hemoglobin concentration. This leads to a reduction in fetal $O_2$ delivery in the human and in sheep to a reduction in both $O_2$ delivery and consumption. This is accompanied by a corresponding decrease in fetal $O_2$ requirements, via reductions in fetal growth rate and body and breathing movements, perhaps in an attempt to match fetal $O_2$ requirements to $O_2$ delivery. If this hypothesis is correct, and there is a causal link between the decrease in fetal oxygen delivery and the reductions in fetal somatic activity, several questions remain. Firstly, what are the mechanisms involved in the decrease in weight normalized uterine and umbilical blood flows, or conversely why don't these flows increase progressively to meet fetal oxygen and nutrient requirements? Secondly, what are the mechanisms that link the decrease of fetal oxygen delivery and blood oxygen tension with the reductions in fetal somatic activity and growth rate? In the sections below, these questions will be addressed.

2.7.1 Uterine and Umbilical Blood Flows

A teleological explanation for the failure of uterine and umbilical blood flow to keep pace with fetal growth may be related to an upper limit of the proportions of cardiac output that the mother and fetus can divert to the placenta because of the need to maintain an adequate blood flow to other organs and tissues. In late gestation, the proportion of maternal cardiac output delivered to the uterus is 12% in the human[65] and 16% in sheep; in the latter species this is double the proportion delivered to the kidneys[130]. The proportion of fetal combined ventricular output delivered to the placenta is 33% in the human[73] and 42% in sheep[31]. Although maternal cardiac output increases during pregnancy, this increase must also serve the increased perfusion requirements of other maternal tissues, such as the kidney, non-placental uterus, extremities, skin and breasts[131]. Moreover, the bulk of the increase in cardiac output occurs in the first half of pregnancy, with little change thereafter[132,133], that is with little change during the period of gestation when fetal weight and thus metabolic demands are increasing rapidly. In terms of fetal combined ventricular output, in humans the weight-normalized value does not change during gestation[73]. This also seems to be the case in sheep[31,134], and in this species there is a progressive reduction in the proportion of cardiac output to the placenta and an increase in the proportions delivered to the lungs and gut[134]. In the human fetus, it is very likely that there is an increased proportion of cardiac output delivered to the brain with advancing gestation[135]; moreover, the much larger fetal brain size in the human compared to the sheep is associated with a much higher proportion of cardiac output (40% vs 4%)[73].

Mechanistically, the increase in total maternal and fetal perfusion to the placenta with advancing gestation must involve increased placental vasculogenesis and angiogenesis, coupled with increasing placental weight and maturation. However, it appears that with the 2 species that this article focusses on, human and sheep, the trajectory of placental growth is different. In the human, placental weight and DNA, RNA and protein content increase throughout pregnancy [136-139], whereas in the sheep, the maximum placental weight peaks at about 80 days gestation (term = 147 d), with little change thereafter[140,141]. Nonetheless in both species there is a progressive increase in chorionic villus branching and maturation with advancing gestation[136,141-143].

Over the last ~15 years, there has been increasing interest in the pro- and anti-angiogenic factors involved in placentation. In human studies, the initial focus was on the mRNA and protein levels of several pro- and anti-angiogenic factors in placental tissues obtained at different gestational ages and their specific locations within maternal and fetal placental cell types, determined using *in-situ* hybridization and immunohistochemical techniques[144-146] The factors examined include acidic and basic fibroblast growth factors, vascular endothelial growth factor (VEGF),which exists as several isoforms, and the related placental growth factor (PlGF) and hepatocyte growth factor, as well as their cellular and soluble receptors, including the tyrosine kinase receptors, Fms-like tyrosine kinase receptor (Flt-1) and kinase-inserted domain receptor (KDR). These receptors mediate the actions of both VEGF and PlGF, which act as pro-angiogenic factors. There is also a soluble form of Flt-1 (sFlt-1), which results from alternate splicing of the receptor[147]. It can bind VEGF and sequester it from binding the membrane bound receptor. It thus has anti-angiogenic actions, as does another soluble receptor, soluble endoglin (Seng), derived from alternative splicing of endoglin, a transmembrane receptor for transforming growth factor ß (TGF-ß)[147]. Moreover, VEGF, PlGF, sFlt-1 and Seng, are released from the placenta into maternal blood[148,149]. However, at least at term, free VEGF is not present in maternal blood, whereas measurable levels are present in cord blood. The lack of free VEGF in the maternal circulation seems to be due to the high concentrations of SFlt-1 that binds the circulating VEGF and prevents it from binding to its membrane-bound receptors. There have been numerous studies that have reported that altered concentrations of PlGF, sFlt-1 and Seng are associated with an increased risk of a number of pregnancy complications including preeclampsia[150], fetal growth restriction[151-154] and stillbirth[155-157]. In all cases, the increased risk of these pregnancy complications was associated with a decreased maternal serum concentration of PlGF and/or increased concentrations of sFlt-1 and Seng. *In vitro* studies of placental tissues obtained following delivery from healthy pregnancies, and from singleton pregnancies associated with severe fetal growth restriction, small-for-gestational age fetuses with normal uterine and umbilical artery Doppler, twin pregnancies with discordant fetal growth and twin pregnancies with preeclampsia, have reported increased mRNA and protein levels of sFlt-1 in placental tissues associated with severe fetal growth restriction, from the smaller of the discordant twins and in one twin in preeclamptic pregnancies, but not with small-for-gestational age fetuses with normal Doppler findings[153].

Gu et al [158] have reported that cultured trophoblast cells from preeclamptic pregnancies had increased release into the culture medium of sFlt-1, Seng and PlGF compared with control trophoblasts. Moreover, when the cells were cultured under hypoxic conditions, the levels of Seng were increased in the trophoblast cells obtained from preeclamptic pregnancies, whereas the concentration of PlGF was reduced in the trophoblast cells obtained from both preeclamptic and control pregnancies. Nagamatsu et al[159] measured the mRNA levels of sFlt-1, membrane bound sFlt-1, VEGF and PlGF in primary cultures of cytotrophoblasts, villous fibroblasts and umbilical vein endothelial cells and the concentrations of sFlt-1, VEGF and PlGF in the culture media in samples obtained from first and second trimesters (5-20 weeks' gestation) placenta following elective pregnancy termination. In the cultures exposed to low $O_2$ concentrations, the sFlt-1 mRNA levels and concentrations in the culture medium were significantly increased in cytotrophoblasts cultures (Figure 19A), but not in villous fibroblast or umbilical vein endothelial cells cultures. The mRNA levels of membrane bound Flt-1 also increased with low oxygen in cytotrophoblasts cells and umbilical vein endothelial cells but not in villous fibroblasts. In contrast, the PlGF concentration in the cytotrophoblast culture media was reduced with decreasing $O_2$ concentrations (Figure 19B). Nevo et al[160] reported that the mRNA levels of sFlt-1 were high in placenta

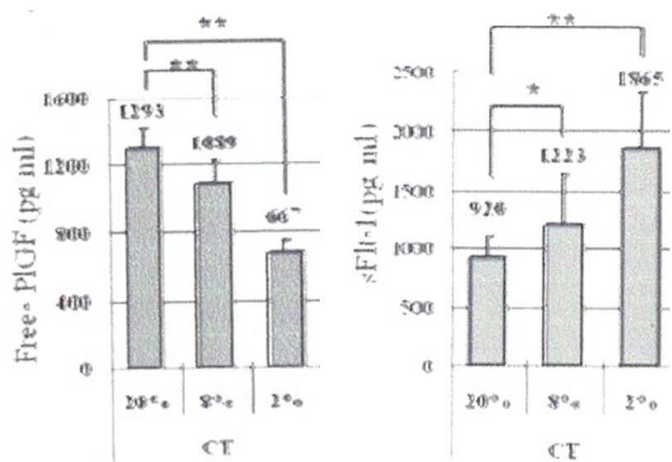

Figure 19. The concentrations of SFlt-1 (A) and PlGF (B) in the media of primary cultures of first trimester human placental cytotrophoblast cells exposed to 20%, 8% and 2% $O_2$. From Nagamatsu et al[159]

samples at 5-8 weeks gestation and decreased at 10-12 and 13-38 weeks, indicating that the surge in placental oxygenation that occurs at 10-12 weeks[161,162] results in a down-regulation of sFlt-1 in the placenta. Elevated sFlt-1 levels were also found in placental samples collected at term from patients with severe early onset preeclampsia and from high altitude pregnancies. In addition, in chorionic villous explant cultures, pharmacological stabilization of the levels of the transcription factor, hypoxia inducible factor (HIF) 1α, enhanced sFlt-1 levels under normoxic conditions. In contrast, antisense oligonucleotide knockdown of HIF-1α decreased sFlt-1 levels under hypoxic conditions. This suggests that increased HIF-1α expression is involved in the elevated sFlt-1 concentration in hypoxic conditions. Moreover, gene microarray results from placenta obtained from patients with early onset preeclampsia, from placenta obtained from high altitude pregnancies and from first trimester villous implants cultured under low oxygen conditions showed similar results suggesting that both early onset preeclampsia and high altitude pregnancy are associated with placental and likely fetal hypoxia[163].

What is more relevant to the current discussion are the longitudinal changes in maternal circulating concentrations of pro and anti-angiogenic factors in normal pregnancies. Several studies have measured one or more of these factors during the course of normal gestation [149,154,157,164]. The results from these studies are similar; maternal plasma PlGF levels increase progressively from early gestation to ~30 weeks gestation and then decline progressively (Figure 20). In contrast, the levels of sFlt-1 and sEng are relatively constant until 25-27 weeks after which they increase progressively to a maximum at term. Thus, the ratio of PlGF/(Seng + sFlt-1), which is a measure of angiogenic activity, increases until ~27 weeks then declines and is at a minimum at term (Figure 20)[165]. These data suggest that placental angiogenesis may decrease in late gestation and this could be a factor in the failure of uterine and umbilical blood flows to keep pace with fetal growth. The stimulus for increased sFlt-1 and Seng levels

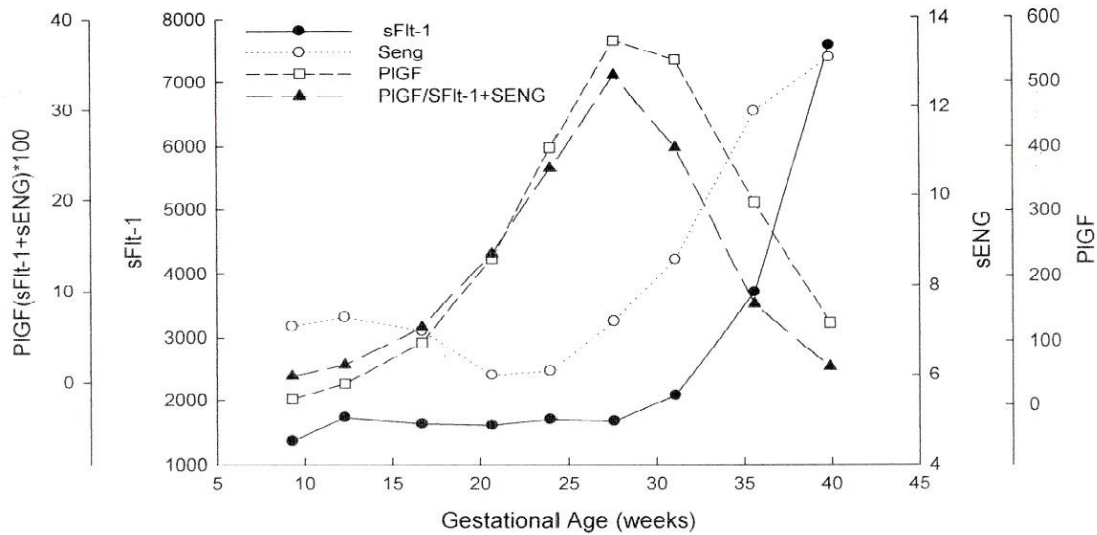

Figure 20. Plots of maternal blood concentrations of placental growth factor (PlGF), soluble Flt-1 (sFlit-1) and soluble endoglin (Seng) during human gestation. Also plotted is the ratio of PlGF concentration divided by the sum of the concentrations of sFlt-1 and Seng. Data plotted from Romero et al[149]

and decreased PlGF in late gestation could be the progressive decrease in fetal vascular and intervillous blood $Po_2$ with advancing gestation (see above).

There are fewer studies of placental pro and anti-angiogenic factors in sheep compared to the human. Cheung and Brace[166] measured the mRNA concentration of VEGF isoforms in the fetal component of the placentome (cotyledon) as a function of gestational age. The major VEGF isoform expressed in the cotyledon was $VEGF_{164}$ and the levels are increased with the main rise occurring from 100 days gestation to term. There was also a linear increase in KDR receptor mRNA with gestational age. A larger number of animals were studied by Borowicz et al[167] who used morphometric techniques to determine the pattern of vascular development in maternal and fetal components of the placentome associated with measurement of the placental mRNA levels of VEGF and its receptors and a number of other angiogenic factors. The vascularity measures all increased in the maternal and fetal placental components with gestational age, except for mean capillary area, which decreased in the cotyledon. VEGF mRNA in the cotyledon increased from 50 to 90 days gestation then was unchanged to term. In contrast in the caruncle (the maternal component of the placenta), VEGF mRNA decreased significantly from 130 to 140 days gestation. There was no significant change in the mRNA concentration of the KDR receptor in the cotyledon, whereas in the caruncle the levels peaked at 110 days and then declined. The Flt-1 mRNA increased in the caruncle until 90 days gestation and was then stable, while in the cotyledon there was a modest increase with gestational age. These results suggest the placental angiogenesis could decrease in late gestation due to a decrease in both VEGF mRNA levels and the reKDR mRNA levels at this time. Grazul-Bilska et al[168] conducted similar analyses in non-pregnant ewes and in pregnant animals from 14 to 30 days gestation. PlGF mRNA levels increased progressively over this time as did FLT-1 mRNA, whereas KDR mRNA peaked at 20 days with little change thereafter. Vonnahme at al[169] measured plasma nitric oxide and VEGF in non-pregnant and postpartum ewes and in pregnant ewes carrying singleton or multiple fetuses between 27 days gestation and term. Plasma VEGF was detected in all groups and during pregnancy the concentrations in multiple pregnancies were either higher or lower than the value in singleton pregnancies at different stages of pregnancy. In multiple pregnancies, the VEGF level fell from the non-pregnant value between 27 and 69 days and at 130-139 days. At no point in pregnancy did the VEGF concentration exceed the non-pregnant value and the post-partum value was the same as the non-pregnant value. In contrast, the plasma NO concentration increased progressively during pregnancy, except between 70 and 89 days. Peak concentrations occurred at 130-139 days and there was no further increase at term, although the concentration fell markedly following delivery. In late gestation, the NO concentration in multiple pregnancies was higher than in the singleton pregnancies. The VEGF results are different from those reported in human pregnancy, since as reported above, free VEGF cannot be detected in maternal plasma at delivery[165]. It is not clear whether any of the other pro- and anti-angiogenic factors that have been identified in human maternal plasma are present in pregnant sheep plasma, although Vonnahme et al[169] report that they could not detect any factor that could bind VEGF in plasma samples for pregnant pigs and sheep, suggesting a lack of sFlt-1 in these species. It may be that the epitheliochorial placentae in these species do not release placentally derived pro- and anti-angiogenic factors into the maternal circulation. If this is the case, the VEGF detected in maternal sheep plasma may come from non-placental sources. More work needs to be done in the sheep and other species with the same placental type to determine, for example, whether cultured trophoblasts release VEGF, PlGF, and the soluble receptors into the medium and whether these factors are present in maternal plasma.

To summarize, there is abundant data from humans suggesting a decrease in placental angiogenesis in late gestation. In sheep, there are conflicting data, but some results also indicate a decrease in placental angiogenesis.

2.7.2 Fetal Motility

In late gestation, the fetuses of precocial species (those mature at birth) such as the human and sheep, are asleep for ~95% of the time. They cycle between rapid eye movement (REM) and quiet sleep and also have brief periods of an aroused state associated with vigorous breathing and body movements and a transient decrease in vascular $Po_2$[170]. This aroused state does not represent wakefulness; wakefulness is inhibited by several compounds released from the placenta and also produced in the fetal brain and also by the lack of a temperature gradient between the fetal surface and core[171,172]. The inhibitory compounds include metabolites of progesterone (neurosteroids) including pregnanolone and allopregnanolone, adenosine and prostaglandin $E_2$. Studies in fetal lambs have demonstrated that neurosteroids inhibit fetal breathing movements and arousal, and increase the incidence of quiet sleep, via interaction with brain $GABA_A$ receptors[173-177]. Conversely, inhibition of progesterone and neurosteroid synthesis increases fetal arousal[173,178].

Nguyen et al[179] measured the mRNA levels of cytochrome P450 side-chain cleavage enzyme, which catalyzes the conversion of cholesterol to pregnanolone, and of 5α-reductase, which catalyzes the conversion of progesterone to 5α-dihydroprogesterone, in various brain regions and as a function of gestational age, and also determined the concentrations of allopregnanolone, pregnenolone and progesterone in the same brain regions and in fetal plasma. The mRNA levels on the steroidogenic enzymes were highest in the midbrain, medulla and pons. They were detected at the earliest gestational age examined (72 d) and increased to term. Following birth, the levels were maintained or increased further. In contrast, the concentrations of the steroids in brain and plasma were highest in late gestation and decreased markedly at birth (Figure 21).

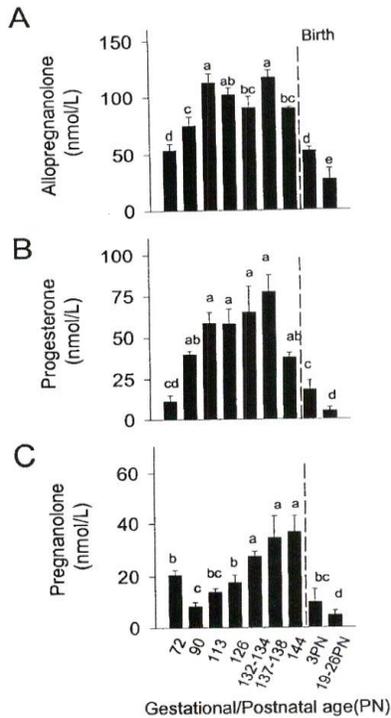

Figure 21. Mean ± SE plasma concentrations of allopregnanolone, progesterone and pregnanolone in fetal lambs as a function of gestational age. From Nguyen et al[179].

The high neurosteroid levels in late gestation are consistent with the decreased motility at this time. Data on the concentrations of neurosteroids in the human fetus as function of gestational age are not available. However, the human maternal plasma concentrations of allopregnanolone increase in parallel with progesterone during gestation[180], and allopregnanolone and pregnenolone were detected in fetal serum samples collected via cordocentesis at 18-23 weeks' gestation[181]. At delivery, the neurosteroids are present in both maternal and cord blood. The concentrations of progesterone are higher in the umbilical vein than in maternal blood and also higher than concentrations in the umbilical artery [182], suggesting release of the steroid from the placenta into the fetal circulation. Pregnenolone concentrations are also higher in umbilical than maternal plasma; however for the other neurosteroids, there are no significant differences between the maternal and fetal unconjugated concentrations[182-184]. Moreover, the allopregnanolone concentrations at delivery are higher in both maternal and fetal plasma compared with mid-pregnancy[181,183]. Following delivery, the maternal concentration of the neurosteroids falls markedly[184]. Thus, in both the fetal lamb and human fetus, there are increasing concentrations in late gestation, which would suppress arousal and motility, while the postnatal decrease in the levels would promote arousal of the neonate.

In the fetal lamb, the circulating concentration of $PGE_2$ is high, due to placental production and release into the fetal circulation[185], and a low percentage of fetal cardiac output that is delivered to the lungs, the major site of prostaglandin clearance. Wallen et al[186] infused meclofenamate intravenously to fetal lambs at 129-136 day's gestation to inhibit prostaglandin synthesis. This dramatically reduced circulating $PGE_2$ levels and was associated with a large increase in the incidence of fetal breathing activity. Co-administration of increasing doses of $PGE_2$ along with meclofenamate increased circulating $PGE_2$ levels and this resulted in a progressive decrease in the incidence of fetal breathing. This provides strong evidence for a role of $PGE_2$ in the episodic pattern of fetal breathing. Beginning about 121-130 days gestation, there is a progressive rise in fetal circulating $PGE_2$ concentration (Figure 22A), which parallels the prepartum rise in fetal plasma cortisol concentration[187]. As illustrated in Figure 22B, these changes are also associated with a progressive increase in the capacity of microsomes isolated from placental cotyledons to metabolize arachidonic acid, which begins after 100 days gestation and leads to the production of $PGE_2$ and $PGF_{2\alpha}$ [188]. Challis et al[189] have summarized the data that indicates that it is the prepartum cortisol rise that increases the expression of prostaglandin synthase II, which results in the increase in $PGE_2$ synthesis and its release into the fetal circulation. Patrick et al induced labor in sheep via intravenous infusion of ACTH and found an inverse relationship between the fetal plasma $PGE_2$ concentration and the incidence of fetal breathing[190]. Wallen et al[191] continuously infused intravenous meclofenamate to fetal lambs over the last 6 to 13 days before delivery, which is a time when the plasma $PGE_2$ concentration is increasing, and the incidence of fetal breathing is decreasing. Meclofenamate decreased the $PGE_2$ concentration and increased the amount of fetal breathing, except during labor, when the $PGE_2$ concentration increased and the incidence of fetal breathing decreased in spite of the continuation of the meclofenamate infusion. Overall, these results indicate that the prepartum rise in plasma $PGE_2$ concentration in fetal lambs is involved in the progressive decrease in fetal breathing at this time and that the elevation in $PGE_2$ level is due to the prepartum rise in fetal plasma cortisol concentration. Given that the prepartum rise in fetal cortisol level in sheep is linked to the onset of parturition[189], this links the timing of the decrease in fetal breathing to the end of gestation.

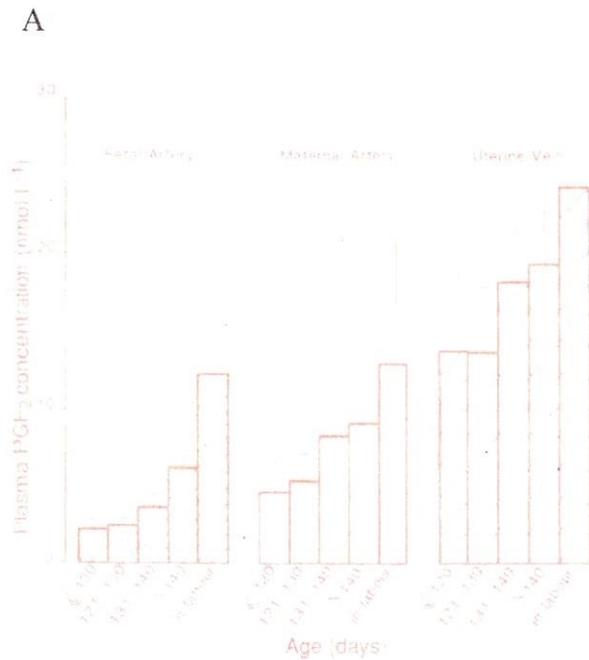

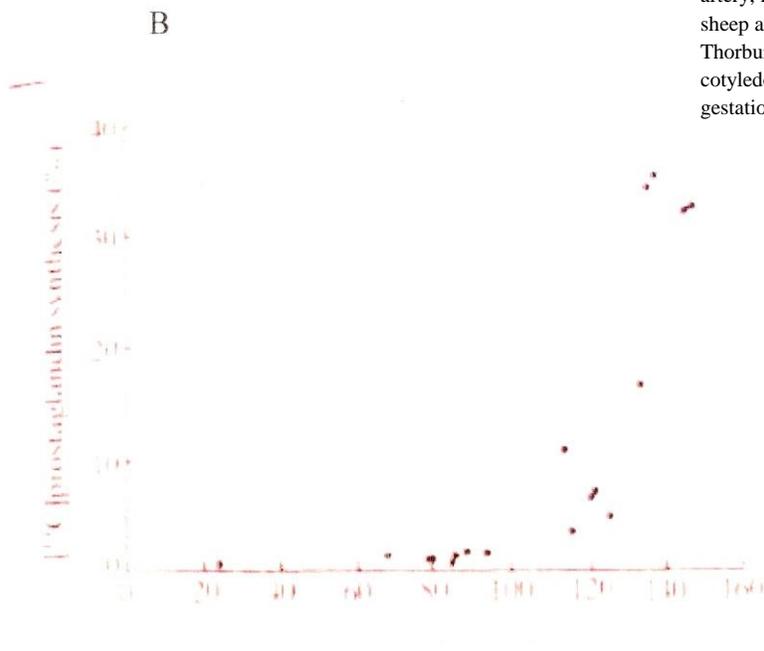

Figure 22. A) Plasma concentration of $PGE_2$ in a fetal artery, maternal artery and uterine vein in pregnant sheep as a function of gestational age. From Thorburn[187]. B) Synthesis rate of $PGE_2$ in placental cotyledonary microsomes are a function of gestational age. From Rice et al[188]

The mechanisms underlying the maturation of the fetal hypothalamic–pituitary–adrenal (HPA) axis are complex[189]. One mechanism that has been investigated in fetal lambs involves the transient decreases in fetal vascular $Po_2$ that occur with episodic fetal 2activity and pre-labor uterine contractions (contractures)[90,91] (Figure 23). These decreases are of the order of 4-5 mm Hg and are due to a transient decrease in uterine blood flow during uterine contractures[192] and to an increase in fetal $O_2$ consumption associated with fetal breathing and body movements[18,19,21]. Acute fetal hypoxemia increases fetal plasma ACTH and cortisol concentrations, and the threshold for these effects is between 5 and 8 mm Hg at 125-

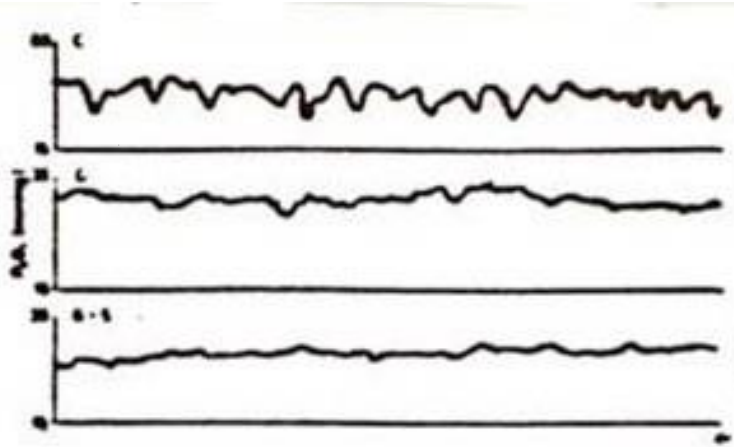

Figure 23 (A) Continuous recordings of arterial $Po_2$ in a fetal lamb for 4 h periods. The upper trace illustrates the control condition, the middle trace after i.v. injection of a neuromuscular blocking agent and the lower trace after i.v. administration of salbutamol (a $\beta_2$ adrenergic receptor antagonist). Modified from Harding et al[91].

129 days gestation[193]. When transient uterine contractions, similar to those which occur normally in pregnant sheep during gestation, were caused by maternal intravenous injection of a small dose of oxytocin, this resulted in a transient rise in intrauterine pressure accompanied by a fall in arterial $Po_2$ and rise in ACTH (Figure 24)[194]. Thus, in late gestation in sheep, transient reductions in vascular $Po_2$ may result in pulsatile ACTH release and this may be involved in the maturation of the adrenal cortisol response to ACTH that contributes to the prepartum cortisol surge[189]. Whether the progressive decrease in fetal oxygenation with advancing gestation is also involved in this mechanism remains to be determined.

There do not appear to be any data on human fetal $PGE_2$ concentrations as a function of gestational age. However, following delivery, cord blood $PGE_2$ concentrations are high and the concentration in umbilical venous blood is higher than in umbilical arterial blood, suggesting release of $PGE_2$ from the placenta into the fetal circulation[195,196]. Moreover, glucocorticoids increase expression of prostaglandin synthase II in human amnion cell and trophoblast cell preparations, as in the sheep, and in addition, corticotrophin-releasing hormone (CRH), which is produced by the placenta, and several cytokines also increase prostaglandin synthase II expression[189]. There is also a prepartum cortisol rise in the human fetus, as there is in sheep and other species[197]. In addition, when controlled release $PGE_2$ vaginal pessaries were inserted into term pregnant women for cervical ripening, the number of fetuses with breathing movements decreased from 59% of the subjects at the time of pessary insertion to 6% and 0% at 6 and 12 h afterwards, respectively, associated with rise in the maternal plasma concentration of 13,14-dihydro-15-keto-$PGE_2$, a stable $PGE_2$ metabolite, from 211±71 to 451±147, and 446±62 pg/ml, respectively[198]. These rather limited data suggest that in the human fetus, plasma $PGE_2$ concentration may rise in late gestation and contribute to the inhibition of breathing movements as in the sheep. However, it is not yet clear if vascular $Po_2$ in the human fetus fluctuates to the same extent as in the fetal lamb. Antepartum uterine contractions in the form of Braxton-Hicks contractions occur in human pregnancy; however, the mean duration of Braxton-Hicks contractions in human pregnancy is 114 sec (1.9 minutes) and only 25% have durations from 2 to 5 minutes[199], compared to the mean duration of contractures in sheep of 6.7 min[200]. Non-invasive measurement of blood $O_2$ saturation in the human fetus using near infrared spectroscopy, as has been done is sheep[201], could address this question.

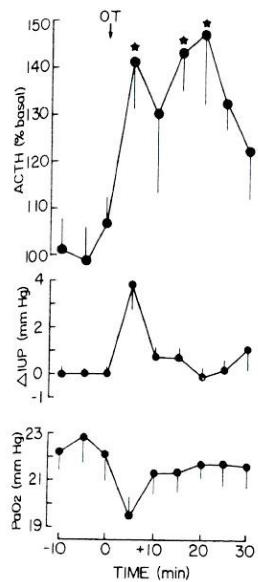

Figure 24. Changes in fetal lamb plasma ACTH concentration, intrauterine pressure (IUP) and arterial Po2 before and after iv. Injection of oxytocin to the ewe (at time 0). From Lye et al[189].

Adenosine is critical for $O_2$ sensing and in the cardiovascular, respiratory, metabolic and endocrine response to hypoxia[202]. It is produced from adenosine monophosphate (AMP) by 5'a-nucleotidases and intracellular adenosine levels rise with hypoxia because of an increase in the AMP/ATP ratio. Once formed, it can diffuse from the cell into the extracellular space[202]. In the fetal lamb, acute hypoxia results in the inhibition of breathing movements and rapid eye movements, and these effects are abolished by co-administration of an adenosine $A_{2A}$ receptor blocker[203]. At ~128 d gestation, acute hypoxia also increases plasma adenosine concentration (1.1±0.2 to 2.0±0.4 μM) and the concentration is inversely related to arterial $O_2$ content[204]. During hypoxia, the degree of inhibition of fetal breathing is linearly related to brain extracellular adenosine concentration[205]. Intravenous infusion of adenosine to the fetal lamb to increase plasma adenosine concentration from 1.1±0.2 μM to 2.8±0.5 μM resulted in marked reductions in rapid eye movements and breathing movements, i.e., similar to the effects of acute hypoxia and a lesser reduction in the amount of low voltage electrocortical activity[204]. Fetal plasma adenosine concentration at 135-140 d gestation averaged 1.9±-0.3 μM[206] and this value is significantly higher (p<0.02) than the value at 128 d (1.1±g0.2 μM) given above. Collectively, these results suggest the rise in plasma adenosine concentration in late gestation would be sufficient to contribute to the reduction in fetal breathing that occurs over this time. It could also contribute to the reduction in REM sleep that occurs in late gestation[207]. This would also decrease fetal oxygen demands as cerebral $O_2$ consumption in the fetal lamb is higher in REM sleep that in quiet sleep[208].

In 33 normal human pregnancies at term, the plasma adenosine concentration in umbilical cord blood averaged 0.23±0.01 μM[209]. This is considerably lower than the arterial adenosine concentration in fetal lambs at 135-140 d gestation (1.9 μm) mentioned above. In pregnancies complicated by maternal diabetes, the adenosine concentration in umbilical venous blood is higher (~0.7 μM) than in normal pregnancies and also higher than in umbilical arterial blood[210], suggesting release of adenosine from the placenta into the fetal circulation under this condition. Adenosine relaxes human umbilical vein rings in vitro and this primarily involves activation of adenosine $A2_A$ receptors[211], the same receptor involved in adenosine's effects on REM sleep and breathing in fetal lambs (see above). Given the apparent universal involvement of adenosine in cardiovascular, metabolic and endocrine control, it seems likely that adenosine effects in the human fetus are similar to those in the fetal lamb.

### 2.7.3 Fetal Growth

There are abundant data to support the view that in fetal lambs the slowing of growth in late gestation is due to the prepartum cortisol rise. Fowden et al[212] used an implanted crown-rump measuring device to longitudinally monitor growth in normal fetal lambs and in fetal lambs that had been earlier subjected to bilateral adrenalectomy.

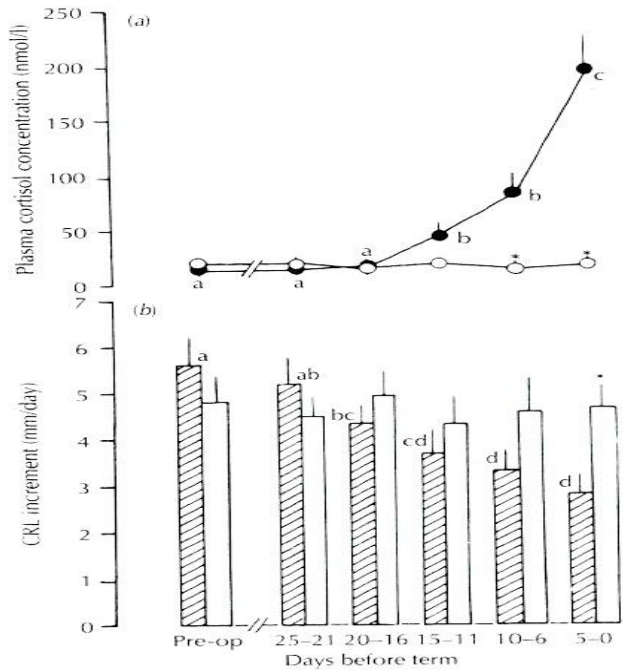

Figure 25. Plasma cortisol concentrations in normal fetal lambs (solid symbols) and in fetuses subject to bilateral adrenalectomy (open symbols) as a function of gestational age, here plotted as days before term. b) The daily increment in fetal crown-rump length in normal fetal lambs (hatched column) and in adrenalectomized fetuses (open columns)[212]

In the control fetuses, the daily increase in crown-rump length began to decrease coincident with the onset of the prepartum cortisol rise, whereas in the adrenalectomized fetuses growth rate was unaltered until term (Figure 25). In contrast in adrenalectomized fetuses, administration of exogenous cortisol reduced fetal growth coincident with the increase in fetal plasma cortisol. A similar experimental approach with normal and adrenalectomized fetal lambs was used to examine the effects of cortisol on hepatic IGF-II mRNA levels[213]. The levels decreased in parallel with the prepartum cortisol, but this did not occur in adrenalectomized fetuses, except when exogenous cortisol was administered. IGF-II is the major IGF in the fetal lamb[214], so cortisol effects on IGF-II expression could be a major mechanism involved in cortisol effects on fetal growth.

As noted above, the prepartum cortisol rise has been observed in a number of mammalian species including the human[197]. In cord blood at term, the IGF-II concentration is considerably higher than the IGF-I concentration[215] and fetal tissue IGF-II expression is more extensive than for IGF-I[216] as in the sheep. Braun et al[217], in a retrospective cohort study, examined the effects of varying maternal doses of betamethasone on perinatal outcomes in singleton pregnancies. Birth weight was significantly lower in pregnancies associated with betamethasone administration compared to controls and the difference in birth weight increased with increasing dose of the steroid. This was also the case with fetal weight gain between 28 week's gestation and term in both males and females. Several randomized trials have also shown reduction in birth weight following repeated doses of antenatal corticosteroids administered to women with threatened preterm birth[218]. A randomized trial by Murphy et al[219] involving 1858 pregnant women at risk of preterm birth showed that infants whose mothers received multiple doses of corticosteroids weighed less at birth than those exposed to placebo (2216 g vs 2330 g), were shorter (44.5 cm vs 45.4 cm), and had a smaller head circumference (31.1 cm vs 31.7 cm). Thus, maternal administration of a synthetic glucocorticoid in late gestation can reduce fetal growth rate in the human, raising the possibility that the fetal prepartum surge in endogenous cortisol could elicit the same effect.

2.8 Stillbirth in Animals and Experimental Methods to Prolong Gestation

There are less data on stillbirth rates in animal species than in humans. Dawes[11] provides estimates of perinatal loss rates in rats (1-2%), pigs (4-12%) and sheep (2-6%) which are much higher than the human rate in developed countries. In addition, the Dawes estimate includes both stillbirth and neonatal death rates. Recently, we determined the stillbirth and neonatal death rates over an 11-year period (1990-2000) in the sheep breeding flock that used to exist at the University of British Columbia using the breeding records for the flock. The overall rate was 24.5/1000 in the 599 deliveries that occurred, while the neonatal death rate was 13.5/1000. The stillbirth rates in singleton, twins and triplets were 6.2/1000, 37.5/1000 and 22.7/1000, respectively. The stillbirth rate in twins was significantly greater than in singletons. It was not possible to test triplets because of the small number of triplets. These data suggest that multiple pregnancy may also be a risk factor for stillbirth in sheep as it is in the human. These animal data indicate that stillbirth is not a problem restricted to the human, but likely occurs in all eutherian mammals.

In pregnant sheep, several experimental approaches have been used to prolong gestation. Bengtsson and Schofield[220] employed daily intramuscular injections of 80 or 160 mg progesterone or 25 or 40 mg medroxyprogesterone acetate for 7 days prior to expected date of delivery or for a maximum of 14 days. Gestation was prolonged for 10±3 days. However, the stillbirth rate was 96%; the only fetuses that were born alive were in pregnancies extended for only a few days. Similar results of a high fetal mortality have been obtained in pregnant rabbits[221]. Liggins et al[222] electrocoagulated the pituitary gland in fetal lambs at ~90 days gestation (6 weeks prior to term). This extended gestation by 21±2 days). This provided some of the first evidence for the involvement of the fetal hypothalmo-pituitary-adrenal (HPA) axis in the initiation of parturition in sheep[189] However in contrast to the results of maternal progestin injections, the stillbirth rate was 0%. An explanation for the dramatically different results of maternal progestin administration and fetal hypophysectomy resides in the fact that the latter procedure disrupts not only the fetal HPA axis but also the actions of the other trophic hormones released from the anterior pituitary. The most important of these hormones is thyroid stimulating hormone. Hypophysectomized fetal lambs have very low circulating concentrations of thyroid hormones[223] and thyroid hormones are a major determinant of fetal metabolic rate. Fowden and Silver[224], in studies of control and thyroidectomized fetal lambs with and without exogenous thyroxine, found that there is a statistically significant linear relationship between plasma thyroxine concentration and the rate of fetal $O_2$ consumption. Fetal $O_2$ consumption in the thyroidectomized fetuses was 33% lower than in the control fetuses, which in greater than the percentage fall in fetal O2 consumption that occurs between 75 and 132d gestation in pregnant sheep (24.1%)[99]. Thus, the hypophysectomized fetal lamb, because of the much lower metabolic rate, can tolerate the further reduction in fetal $O_2$ delivery that occurs with prolonged gestation, whereas maternal progestin administration inhibits myometrial contractions with no direct effect on the fetus. Gestation can also be prolonged (by 10±2 days) in sheep by fetal bilateral adrenalectomy[225,226]. However, the incidence of stillbirth was high (`85%), which is similar to the results of maternal progesterone administration. In twin pregnancies, fetal weight was greater in the adrenalectomized fetus compared to the unoperated control lamb[226], which is consistent with lack of the antenatal cortisol rise in the former fetuses not inhibiting fetal growth[212]. Collectively the results support the hypothesis that in sheep and rabbits at least prolongation of gestation beyond normal term results in the maternal-placental-fetal supply line becoming increasingly unable to meet the fetal requirements for $O_2$.

3.0 Summary and Conclusions

The following points have been made. Fetuses receive adequate amount of oxygen to meet maintenance functions. However, the vascular $Po_2$ in the fetus is much lower than after birth so that the fetus is much lower to the critical vascular $Po_2$, below which diffusion of oxygen from blood to tissue is impaired. Moreover, there is evidence that that fetal lamb cannot survive when vascular Po2 is reduced to ~ 10 mm Hg. During gestation, there is a failure of uterine and umbilical blood flows to keep pace with fetal growth (Figure 26). This leads to progressive decreases in fetal vascular $Po_2$, $O_2$ saturation and content and pH and an increase in $Pco_2$. The fall in weight-normalized blood flow coupled with the decline in umbilical venous $O_2$ content decreases fetal $O_2$ delivery and in the sheep and horse this is accompanied by a decrease in fetal $O_2$ consumption.

This reduced $O_2$ consumption is achieved at least in part by a progressive decrease in fetal motility with advancing gestation and a decline in fetal growth rate in late gestation. The failure of placental blood flows to keep pace with fetal growth may simply be due to an inability of the maternal and fetal cardiovascular systems to supply the placenta above a certain limit. Mechanistically, it may involve the GA-related changes in the pro- and anti-angiogenic factors involved in placentation, which in pregnant women are released into the maternal circulation and can be measured. In normal human pregnancy, the ratio of placental growth factor (PlGF) to soluble endoglin concentrations, which is a measure of angiogenic activity, increases until ~27 weeks GA and then declines progressively until term. *In vitro* studies of cultured trophoblasts have indicated that hypoxia decreases PlGF and increases sFlt-1 release, suggesting that the progressive fall in fetal oxygenation with advancing GA may be involved in the change in pro- and anti-angiogenic factors. The decrease in fetal motility with advancing GA may be due to the increase in fetal plasma $PGE_2$ concentration, which begins at ~120 days secondary to the prepartum rise in fetal cortisol. Fetal breathing and REM sleep are decreased by adenosine and the plasma adenosine concentration increases in late gestation in fetal lambs, perhaps in response to the progressive decrease in vascular $Po_2$. In addition, the fetal plasma levels of neurosteroids, which suppress fetal arousal, increase progressively with advancing GA. The prepartum cortisol rise also inhibits fetal growth. This cortisol rise is in part due to increased adrenal sensitivity to adrenocorticotrophic hormone (ACTH). It has been suggested that this maturation is in part due to the transient episodes of fetal hypoxemia that result from antepartum uterine contractions and episodic fetal activity. It could also involve the GA-related progressive fall in fetal vascular $Po_2$. Finally, the studies involving prolongation of pregnancy by maternal progestin administration in sheep and rabbits and adrenalectomy in sheep indicate the gestation can be extended for only a few days beyond term before the rate of stillbirth rises dramatically.

In a normal pregnancy, these mechanisms operate effectively to maintain an appropriate balance between fetal oxygen consumption and delivery, and the fetus is born in good condition at a normal birth weight. However, with pregnancy complications that either further reduce $O_2$ delivery (e.g. preeclampsia, post-term pregnancy) or increase fetal $O_2$ demands (e.g. multiple pregnancy, maternal diabetes), the mismatch between $O_2$ delivery and consumption may worsen leading to fetal growth restriction, hypoxic organ damage or stillbirth. In addition, the fact that gestational age-related decreases in weight normalized uterine and umbilical blood flows, fetal $O_2$ delivery, motility and growth occur in at least 2 mammalian species (humans and sheep), which are not closely related, and have markedly different placentae and gestation lengths, suggests that the length of gestation in different mammalian species may be determined in part by how well fetal oxygen delivery can be maintained with advancing gestation.

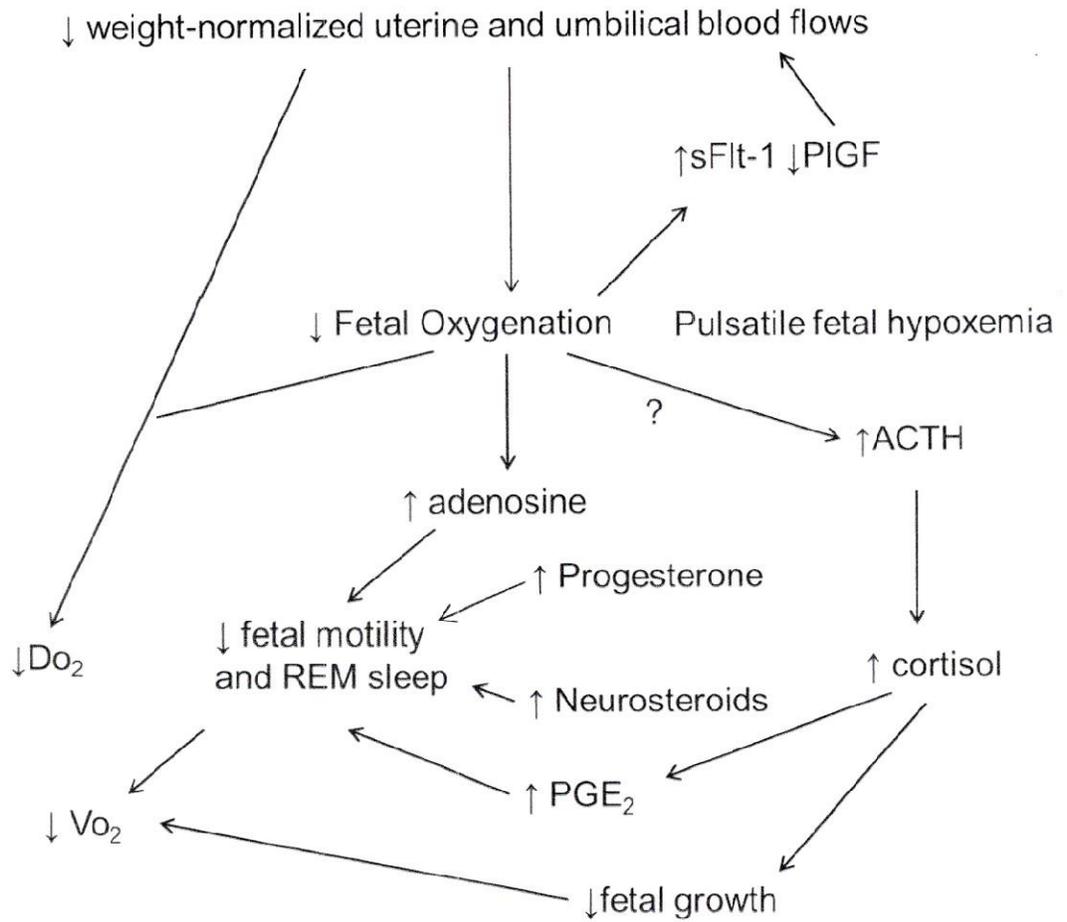

Figure 26. Summary of the postulated mechanism to explain the increased risk of stillbirth, fetal growth restriction and other adverse perinatal outcomes with advancing gestational age.

Although the data presented in this paper support the hypothesis posed, some questions remain. In human pregnancy, fetal vascular and intervillous $Po_2$ values begin to decrease by 16-18 week's gestation[80,81]. At this age, the fetal is small and its metabolic demands low, so that the ratio between uterine $O_2$ delivery and fetal $O_2$ consumption must be high. It may be that at this time, it is the placenta that imposes the major demand upon uterine $O_2$ delivery, since placental $O_2$ consumption is much higher than that of the fetus.[5] In the third trimester it may be the combined $O_2$ demands of the placenta and fetus that become the determining factor. Longitudinal measurements of uterine and umbilical blood flows, $O_2$ delivery and utero-placental and fetal $O_2$ consumption, which is feasible in pregnant sheep[97], could address this issue.

In closing, as noted previously, our interest in the changes in fetal oxygenation with gestation was prompted by the observation that using the fetuses-at-risk approach in the analysis of population data on pregnancy outcome there is an increasing risk of fetal growth restriction, stillbirth and other adverse perinatal outcomes with increasing gestational age[52]. This review provides a biological explanation for this observation, which suggests that the fetuses-at-risk approach for analyzing pregnancy outcome data in relation gestational age is a useful approach.

Acknowledgements: We thank M. Paul Willing, UBC Farm Manager, for providing us with the sheep breeding records.

Conflicts of Interest: The authors declare that they have no conflicts of interest in relation to this manuscript.

27. Itskovitz J, Goetzman BW, Roman C, Rudolph AM. Effects of fetal-maternal exchange transfusion on fetal oxygenation and blood flow distribution. *Am J Physiol*. 1984;247:H655-H660.

28. Reed L, Meijboom EJ, Sahn DJ, Scagnelli SA, Valdes-Cruz LM, Shenker L. Cardiac doppler flow velocities in human fetuses. *Circulation*. 1986;73:41-46.

29. Rudolph AM, Heymann MA. The circulation of the fetus in utero. methods for studying distribution of blood flow, cardiac output and organ blood flow. *Circ Res*. 1967;21:163-184.

30. Tan W, Riggs KW, Thies RL, Rurak DW. Use of an automated fluorescent microsphere method to measure regional blood flow in the fetal lamb. *Can J Physiol Pharmacol*. 1997;75(8):959-968.

31. Anderson DF, Bissonnette JM, Faber JJ, Thornburg KL. Central shunt flows and pressures in the mature fetal lamb. *Am J Physiol*. 1981;241(1):H60-6.

32. Mielka G, Benda N. Cardiac output and central distribution of blood flow in the human fetus. *Circulation*. 2009;73:1162-1168.

33. Hales JR. Radioactive microsphere measurement of cardiac output and regional tissue blood flow in the sheep. *Plugers Archiv*. 1973;344:119-132.

34. Kobe J, Mishra N, Arya VK, Al-Moustadi W, Nates W, Kumar B. Cardiac ouput monioring: Technology and choice. *Ann Card Anaesth*. 2019;22:6-17.

**Figure Legends**

1. $O_2$ dissociation curves for fetal and maternal blood in sheep. The lines intersecting the fetal curves denote $Po_2/O_2$ saturation relationships for umbilical venous (UV) umbilical arterial (A) and fetal systemic venous (V) bloods. (Rurak, unpublished data).
2. The relationship between umbilical venous $Po_2$ and umbilical blood flow in several species. Data are from the following sources: cow[227,228], horse[88,228], sheep: Rurak unpublished data, pig[228,229], human [73,80], baboon[230], Rhesus monkey[231].
3. Plots of control and experimental values for arterial $Po_2$, umbilical venous $Po_2$, umbilical blood flow, oxygen delivery, oxygen consumption and arterial lactate concentration after 1.7 h of fetal hypoxemia[39] and 2 h after 40-45% acute hemorrhage[40].
4. Arterial $Po_2$, $Pco_2$, pH and hematocrit (PCV) in a chronically instrumented, spontaneously growth-restricted fetal lamb in late gestation. (Rurak, Unpublished data)
5. The relationship between gestational age and the rate of stillbirth and   for gestational age in human singleton and twin pregnancies[52].
6. Uterine blood flow normalized to fetal weight (mean±SE) as a function of gestational age in human pregnancy. Data plotted from [66].
7. Uterine blood flow as a function of gestation normalized to fetal weight for several mammalian species. Reproduced by the kind permission of Recent Advances in Physiology, from [232].
8. Human umbilical blood flow normalized to fetal weight as a function of gestational age. Plotted from [95].
9. Umbilical blood flow/kg fetal weight in the fetal lamb as a function of gestational age. Each symbol represents a different fetus. D Rurak, unpublished data.
10. Values of umbilical venous $Po_2$ ($UVpO_2$), Pco2 (UVpCO2), pH (UVpH) and total red cell count as a function of gestational age in human fetuses subject to cordocentesis. Adapted by the kind permission of Obstetrics and Gynecology, from [81].
11. Values of fetal arterial Po2, Pco2, pH and $O_2$ saturation as a function of gestational age in 447 blood samples from 108 pregnant ewes from 103 to 146 d gestation[94].

12. Mean±SE values of $O_2$ consumption in fetal lambs in relation to gestational age. Plotted from [99].
13. Regression curves for the relationship between gestational age and the percentage of fetal movements (a), the number of fetal movements (b), the mean duration of movement bouts (c) and the onset to onset interval (d) in 29 human fetuses from normal pregnancies. Reproduced by the kind permission of Pediatric Research, from [107].
14. Mean±SE values for fetal body movements/min (A) and minutes of quiescence (B) in 8 fetal lambs as a function of gestational age[108].
15. Mean growth velocities for head volume (HV), femur diaphase length (FDL), occipitalfrontal diameter (OFD), head circumference (HC), biparietal diameter (BPD) and head circumference (HC) (as a percentage of the size at 40 week's gestation for the human fetus. Reproduced by the kind permission of Archives of Diseases in Childhood Fetal and Neonatal Edition from [124].
16. The mean daily crown rump length increment (upper graph) and maternal plasma glucose concentration between 100 and 143 d gestation in control pregnant sheep (A) and in moderately undernourished (B) and severely undernourished (C) animals. Reproduced by the kind permission of Quarterly Journal of Physiology and Cognate Medical Sciences from [129].
17. Longitudinal ultrasound measurements of abdominal diameter in fetal lambs as a function of gestational age[94].
18. The relationship between the gestational age at which body movements first decreased (movement breakpoint) and the gestational age at which abdominal diameter first decreased (AD breakpoint) in 8 fetal lambs[94].
19. The concentrations of SFlt-1 (A) and PlGF (B) in the media of primary cultures of first trimester human placental cytotrophoblast cells exposed to 20%, 8% and 2% $O_2$. Reproduced by the kind permission of Endocrinology from [159].
20. Plots of maternal blood concentrations of placental growth factor (PlGF), soluble Flt-1 (sFlit-1) and soluble endoglin (Seng) during human gestation. Also plotted is the ratio of PlGF concentration divided by the sum of the concentrations of sFlit-1 and Seng. Data plotted from [149].
21. Mean ± SE plasma concentrations of allopregnanolone, progesterone and pregnanolone in fetal lambs as a function of gestational age. Reproduced by the kind permission of Journal of Maternal, Fetal and Neonatal Medicine from [233].
22. A) Plasma concentration of $PGE_2$ in a fetal artery, maternal artery and uterine vein in pregnant sheep as a function of gestational age. Reproduced by the kind permission Reproduction Fertility and Development from [187]. B) Synthesis rate of $PGE_2$ in placental cotyledonary microsomes are a function of gestational age. Reproduced by the kind permission of Journal of Endocrinology from [188].
23. A) Continuous recordings of arterial $Po_2$ in a fetal lamb for 4 h periods. The upper trace illustrates the control condition, the middle trace after i.v. injection of a neuromuscular blocking agent and the lower trace after i.v. administration of salbutamol (a ß$_2$ adrenergic receptor antagonist). Modified by the kind permission of Journal of Developmental Physiology from [91].

24. Changes in fetal lamb plasma ACTH concentration, intrauterine pressure (IUP) and arterial Po2 before and after iv. Injection of oxytocin to the ewe (at time 0). Reproduced by the kind permission of Journal of Endocrinology from [194].

**25.** a) Plasma cortisol concentrations in normal fetal lambs (solid symbols) and in fetuses subject to bilateral adrenalectomy (open symbols) as a function of gestational age, here plotted as days before term. b) The daily increment in fetal crown-rump length in normal fetal lambs (hatched column) and in adrenalectomized fetuses (open columns). Reproduced by the kind permission of Journal of Endocrinology from [212].

**26.** Summary of the postulated mechanism to explain the increased risk of stillbirth, fetal growth restriction and other adverse perinatal outcomes with advancing gestational age. See text pages 30-31